\documentclass[aps,prd,superscriptaddress,twoside,twocolumn,nofootinbib,%
10pt,showpacs,floatfix]{revtex4-1}

\usepackage{amsmath,amssymb}
\usepackage{graphicx,bm}
\usepackage{ulem}
\usepackage{slashed}
\usepackage{dcolumn}
\usepackage{epsfig}
\usepackage{multirow}

\allowdisplaybreaks

\begin{document}


\title{Further signatures to support the tetraquark mixing framework for the two light-meson nonets}


\author{Hungchong Kim}%
\email{hungchong@kau.ac.kr}
\affiliation{Research Institute of Basic Science, Korea Aerospace University, Goyang, 412-791, Korea}
\affiliation{Center for Extreme Nuclear Matters, Korea University, Seoul 02841, Korea}

\author{K. S. Kim}%
\affiliation{School of Liberal Arts and Science, Korea Aerospace University, Goyang, 412-791, Korea}

\author{Myung-Ki Cheoun}%
\affiliation{Department of Physics, Soongsil University, Seoul 156-743, Korea}

\author{Daisuke Jido}%
\affiliation{Department of Physics, Tokyo Institute of Technology, Meguro, Tokyo 152-8551, Japan}

\author{Makoto Oka}%
\affiliation{Advanced Science Research Center, Japan Atomic Energy Agency, Tokai, Ibaraki, 319-1195 Japan}

\date{\today}


\begin{abstract}

In this work, we investigate additional signatures to support the tetraquark mixing framework
that has been recently proposed as a possible structure for the two nonets, namely
$a_0 (980)$, $K_0^* (800)$, $f_0 (500)$, $f_0 (980)$ in the light nonet,
$a_0 (1450)$, $K_0^* (1430)$, $f_0 (1370)$, $f_0 (1500)$ in the heavy nonet.
First, we advocate that the two nonets form the flavor nonet approximately satisfying the
Gell-Mann--Okubo mass relation.
Then we reexamine the mass ordering generated from the tetraquark nonets and
show that this mass ordering is satisfied by the two nonets although the ordering in the heavy nonet is marginal.
The marginal mass ordering however can be regarded as another signature for tetraquarks because
it can be explained partially by the hyperfine masses calculated from the tetraquark mixing framework.
The tetraquark mixing parameters are found to be independent of isospins giving additional support
for the formation of the flavor nonets.
In addition, we discuss the other approaches like two-quark pictures or meson-meson bound states, and their possible limitations
in explaining the two nonets.
As a peculiar signature distinguished from other approaches, we investigate the fall-apart coupling
strengths into two vector mesons from our tetraquarks.
Coupling strengths into the two-vector modes are found to enhance strongly in the heavy nonet
while they are suppressed in the light nonet.
The coupling ratios, which depend on the isospin channel, are found to be huge around $\sim 15$.
This trend in the two-vector modes, which is opposite to that in the two-pseudoscalar fall-apart modes, can
provide another testing ground for the tetraquark mixing framework.
Some experimental evidences related to the phenomena are discussed particularly from the resonances belonging to the heavy nonet.

\end{abstract}

\maketitle

\section{Introduction}

For the last decade or so, an increasing number of exotic states reported from worldwide high energy facilities
has triggered great excitement in the hadron community
especially because of the possibility that they might be the long-sought multiquark states.
These exotic states include the pioneering resonance,
$X(3872)$~\cite{Belle03,Aubert:2004zr, Choi:2011fc, Aaij:2013zoa} measured in the $B$-meson
decays and other resonances $X(3823)$, $X(3900)$, $X(3940)$, $X(4140)$, $X(4274)$, $X(4500)$,
$X(4700)$~\cite{Bhardwaj:2013rmw,Xiao:2013iha,Abe:2007jna, Aaij:2016iza,Aaij:2016nsc} as well.
Also the pentaquark candidates, $P_c(4380)$, $P_c(4450)$, have been reported in Ref.~\cite{Aaij:2015tga} from $J/\psi p$ channel
in the $\Lambda_b^0 \rightarrow J/\psi K^- p$ decay.
Recently, Ref.~\cite{D0:2016mwd} has reported the observation of $X(5568)$ from the D0 experiment at the Fermilab claiming that
this could be a tetraquark with four different flavors because a molecular state composed
of loosely bound $B_d$ and $K$ mesons is disfavored due to the large mass difference.

Theoretically, studies on tetraquarks in hadron spectroscopy are very diverse
ranging from the light mesons composed by $u,d,s$ quarks, to the heavy mesons involving charm and bottom quarks.
Even for the heavy mesons, the tetraquark investigation is further subdivided
into various sectors such as hidden-charm~\cite{Maiani:2004vq,Kim:2016tys,Anwar:2018sol,Zhao:2014qva}, open-charm~\cite{Kim:2014ywa},
doubly charmed~\cite{Yan:2018gik,Karliner:2017qjm,Hyodo:2017hue,Esposito:2013fma,Eichten:2017ffp} and, triple~\cite{Chen:2016ont} or fully
charmed~\cite{Lloyd:2003yc,Richard:2017vry,Karliner:2016zzc,Esposito:2018cwh}, and the similar states with bottom quarks.
Eventually a unified approach for tetraquarks is anticipated because all the constituent quarks are bound by the color forces
that are in principle independent on quark flavors.

The most popular approach for tetraquarks is the diquark-antidiquark model~\cite{Jaffe77a,Jaffe77b,Jaffe04}
proposed long ago by Jaffe in his exploratory investigation of tetraquarks in the light mesons.
In this approach, the tetraquarks are constructed by combining diquarks
and antidiquarks. Since diquarks and antidiquarks are colored,
the resulting tetraquarks can form tightly bound states by direct color forces.
So the diquark is a sort of building block in constructing tetraquarks.
A common practice is to use the spin-0 diquark with the color and flavor structures of ($\bar{\bm{3}}_c, \bar{\bm{3}}_f$)
because this diquark is most attractive among all the possible
diquarks~\cite{Jaffe:1999ze} if the binding is mainly driven by the color-spin interaction.
Famous candidates are the light nonet consisting of $a_0 (980)$, $K_0^* (800)$, $f_0 (500)$,
$f_0 (980)$~\cite{Jaffe77a,Jaffe77b,Jaffe04,Jaffe:1999ze,MPPR04a,EFG09,Santopinto:2006my,Agaev:2017cfz,Agaev:2018fvz}.

However, in its extension to the tetraquarks containing heavy quarks,
the possible diquarks are not limited to the spin-0 diquark and different diquarks
are often adopted in the construction of tetraquarks.
This is because of a possibility that
the binding within a diquark can be provided by other mechanisms different from the color-spin interaction.
For instance, the color-electric interaction can participate
in holding the two quarks in a diquark~\cite{Karliner:2017qjm}.
Indeed, people are looking for bound states from doubly charmed
($cc\bar{q}\bar{q}$, $q=u,d,s$)~\cite{Yan:2018gik,Karliner:2017qjm,Hyodo:2017hue,Esposito:2013fma,Eichten:2017ffp},
triple ($cc\bar{c}\bar{q}$)\cite{Chen:2016ont} or fully
charmed ($cc\bar{c}\bar{c}$) tetraquarks~\cite{Lloyd:2003yc,Richard:2017vry,Karliner:2016zzc,Esposito:2018cwh} in the diquark-antidiquark approach
even though the heavy diquark, $cc$, cannot be the spin-0
diquark above~\footnote{Since the diquark flavor, $cc$, is symmetric, its spin and color configurations
are restricted either to ($J=1, \bar{\bm{3}}_c$) or
to ($J=0, \bm{6}_c$). Both configurations are repulsive
in color-spin interaction~\cite{Jaffe:1999ze} and they are clearly different from the spin-0 diquark above.}.
Actually, it is not clear whether the tetraquarks containing $cc$ can form a
bound state or not~\cite{Karliner:2017qjm, Hyodo:2017hue}. For example, Ref.~\cite{Richard:2017vry}
suggested that $cc\bar{c}\bar{c}$ and $bb\bar{b}\bar{b}$ are unbound
while the fully heavy tetraquark with different quark flavors, $bc\bar{b}\bar{c}$, is bound.
The latter result can be understood if one assumes the dominance of the color-spin interaction because this interaction
gives an attraction for the diquark with different quark flavors.

Even in the light meson system composed by $u,d,s$ quarks, one can introduce a different diquark,
namely the spin-1 diquark with ($\bm{6}_c, \bar{\bm{3}}_f$),
and construct the second type of tetraquarks~\cite{Kim:2016dfq,Kim:2017yur,Kim:2017yvd} in addition
to the first type constructed from the most common spin-0 diquark discussed above.
The spin-1 diquark can be used as a building block also because
it forms a bound state even though it is less attractive than the spin-0 diquark above~\cite{Jaffe:1999ze}.
But the total binding, which is calculated by summing over pairwise interactions among all the four quarks,
is found to be more negative so the spin-1 diquark must be considered as an important ingredient in the formation of tetraquarks.
The necessity of the spin-1 diquark is also supported by the QCD sum rule calculation for $D_s(2317)$ using
diquark-antidiquark interpolating fields~\cite{Kim:2005gt}. There, it is found that the interpolating field containing
the vector diquarks can describe $D_s (2317)$ equally well as the interpolating field containing the scalar diquarks.

The main aspect of Refs.~\cite{Kim:2016dfq,Kim:2017yur,Kim:2017yvd} is that the two tetraquark types, one type constructed
from the spin-0 diquark and the other type from the spin-1 diquark,
mix strongly through the color-spin interaction. So the physical resonances
can be identified by the eigenstates that diagonalize the hyperfine color-spin interaction.
This tetraquark mixing framework is very promising as a possible structure for the two nonets in the review of
Particle Data Group (PDG)~\cite{PDG16}, $a_0 (980)$, $K_0^* (800)$, $f_0 (500)$, $f_0 (980)$ in the light nonet,
$a_0 (1450)$, $K_0^* (1430)$, $f_0 (1370)$, $f_0 (1500)$ in the heavy nonet.
Indeed, this structure has been tested relatively well in reproducing the mass splittings between the two nonets as well as
the partial decay widths into two pseudoscalar mesons~\cite{Kim:2016dfq,Kim:2017yur,Kim:2017yvd}.

On the other hand, one can try other approaches for the two nonets in PDG.
One immediate approach would be a two-quark picture with orbital angular momentum $\ell=1$.
But, as we will discuss below,
its simple applications do not explain the two nonets
especially in achieving a phenomenological consistency with the mass ordering.
Another approach is the meson-meson bound states separated with a long-range interaction.
The hadronic molecular picture composed by $K\bar{K}$ or $\pi\eta$ has been proposed for
$f_0(980)$, $a_0(980)$~\cite{Janssen:1994wn,Branz:2007xp,Branz:2008ha,Dudek:2016cru}.
This molecular picture is also actively investigated in the heavy meson sectors including the recent
exotic resonances~\cite{Guo:2017jvc,Wang:2017qbe,Hanhart:2017kbu,Swanson:2004pp,Tornqvist:2004qy,Kim:1995bm}.
As hybrid type approaches, there is a two-quark picture with hadronic intermediate states~\cite{Tornqvist:1995kr, vanBeveren:1986ea}
through the unitarized quark model.
This approach has been extended to generate the physical states
belonging to the two nonets dynamically from a single $q\bar{q}$ state in each isospin channel~\cite{Boglione:2002vv,Wolkanowski:2015lsa}.
Other models also exist in the literature like the tetraquarks mixed with a glueball~\cite{Maiani:2006rq},
the $P$-wave $q\bar{q}$ mixed with the four-quark
$qq\bar{q}\bar{q}$ scalar nonet~\cite{Black:1998wt,Black:1999yz}, or the tetraquarks including instantons~\cite{Dorokhov:1993nw}.
Judging from various approaches, the current status on the nature of the two nonets is rather unclear and more studies are necessary
in order to establish a realistic picture for the resonances being considered here.
However, we believe that our tetraquark mixing framework provides relatively a simple picture
and it may be worth pursuing further consequences of this model.

In this work, we investigate additional signatures to support the tetraquark mixing
framework~\cite{Kim:2016dfq,Kim:2017yur,Kim:2017yvd}
as a plausible structure for the two nonets in PDG, the light and heavy nonets.
First, we point out that the two nonets satisfy the Gell-Mann--Okubo mass relation approximately,
which may indicate that the two nonets form the flavor nonets.
Their tetraquark nature will be justified by demonstrating that their mass ordering is consistent
with the tetraquark picture even though the ordering is marginal in the heavy nonet.
We argue however that the marginal mass ordering can be regarded as another supporting evidence
for our tetraquark mixing framework because it can be explained partially by the narrow splitting between
the hyperfine masses calculated from the tetraquark mixing framework.
Further evidence to support the flavor nonets
can be seen from the fact that the mixing parameters in generating the heavy and light nonets
are almost independent of the isospins. (see the subsection~\ref{sec:mixing parameters} below.)

As comparative models, we examine the other approaches for the two nonets, namely
the two-quark picture with orbital angular momentum $\ell=1$ and the meson-meson molecular picture.
It will be pointed out that the two-quark picture can have only one configuration in the $J^P=0^+$ channel,
which is clearly not enough to accommodate the two nonets in PDG.
Its application only to the heavy nonet is not realistic also due to the inconsistency in the mass ordering.
We discuss the meson-meson picture as well and point out that this picture requires the additional multiplets to be found in PDG.
One possible way to distinguish the tetraquark mixing framework from these pictures would be
the fall-apart modes and their peculiar prediction from the light and heavy nonets.

Along this line, we study the fall-apart modes into the two-vector channels from our tetraquark system
by recombining quarks and antiquarks in the wave functions.
The color and spin factors as well as the flavor recombination factor will be calculated in detail.
Because of the tetraquark mixing framework, the coupling strengths to the two vector mesons are found to be strongly enhanced for
the heavy nonet but they are suppressed for the light nonet.  This is in contrast to the fall-apart modes into the two pseudoscalar mesons
whose coupling strengths are enhanced for the light nonet but suppressed for the heavy nonet~\cite{Kim:2017yur,Kim:2017yvd}.

This paper is organized as follows. After a brief review on the tetraquark mixing framework in Sec.~\ref{sec:mixing framework},
we give additional supporting arguments in Sec.~\ref{sec:candidates} in identifying the two nonets as tetraquarks.
In Sec.~\ref{sec:other}, we introduce other models such as two-quark picture or meson-meson molecule picture and
discuss possible problems with them.
We then present our formalism for the fall-apart modes of our tetraquarks into two vector mesons in Sec.~\ref{sec:fall-apart}
and propose the interesting phenomenological
consequences in the fall-apart strengths that differentiate between the light and heavy nonets.
We summarize in Sec.~\ref{sec:summary}.

\section{Review on tetraquark mixing framework}
\label{sec:mixing framework}

To make our presentation self-contained, we begin by a brief review on the tetraquark mixing framework
advocated in Refs.~\cite{Kim:2016dfq,Kim:2017yur,Kim:2017yvd}.
There, we have demonstrated that two tetraquark types are possible in the $J^P=0^+$ channel within
the diquark-antidiquark model.
The first type is constructed by using the spin-0 diquark with color and flavor structures,
($\bar{\bm{3}}_c, \bar{\bm{3}}_f$).  The second type is constructed from the spin-1 diquark with ($\bm{6}_c, \bar{\bm{3}}_f$).
The spin and color configurations of the two tetraquark types are the followings.
\begin{eqnarray}
&&\text{First type}:\nonumber\\
&&~~|J J_{12} J_{34}\rangle =|000\rangle\ ,\\
&&~~|\bm{1}_c \bar{\bm{3}}_c \bm{3}_c\rangle=\frac{1}{\sqrt{12}}  \varepsilon_{abd}^{} \ \varepsilon_{aef}
\Big ( q^b q^d \Big )
\Big ( \bar{q}^e \bar{q}^f \Big )\label{color 1}\ ,\\
&&\text{Second type}:\nonumber\\
&&~~|J J_{12} J_{34}\rangle=|011\rangle\ ,\\
&&~~|\bm{1}_c \bm{6}_c \bar{\bm{6}}_c\rangle=\frac{1}{\sqrt{96}} \Big( q^a q^b+q^b q^a \Big )
\Big (\bar{q}^a \bar{q}^b+\bar{q}^b \bar{q}^a\Big )\label{color 2}\ .
\end{eqnarray}
Here the state specifications are for tetraquark, diquark, and antidiquark successively.
In particular, $J$ denotes the tetraquark spin, $J_{12}$ the diquark spin, $J_{34}$ the antidiquark spin.
In our discussion below, the two tetraquark types are denoted mostly by their spin configurations,
$|000\rangle$, $|011\rangle$ unless explicit specification of the color configuration is necessary.

\begin{figure}[t]
\centering
\epsfig{file=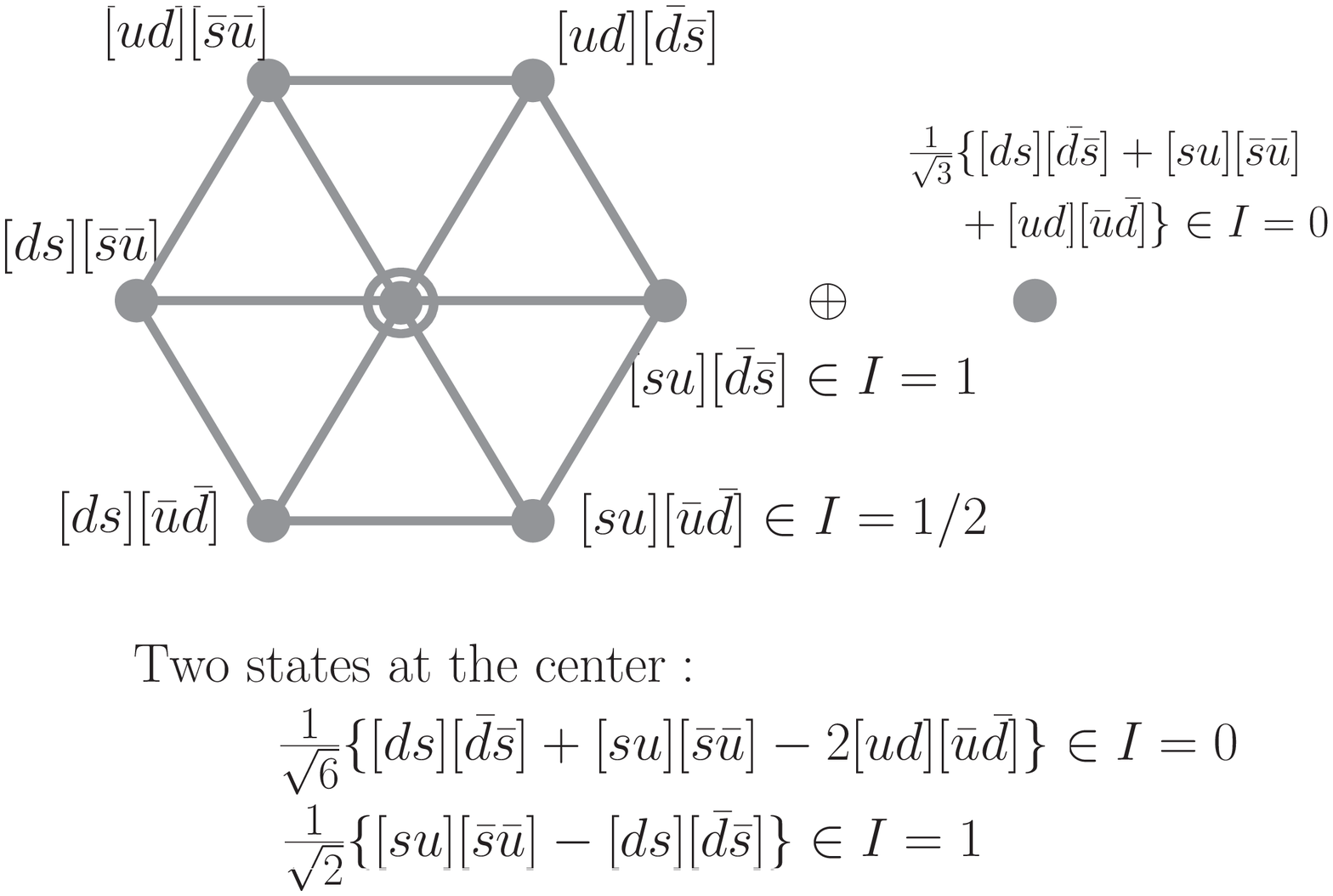, width=0.95\hsize}
\caption{Weight diagram for the tetraquark nonet with explicit flavor wave functions.
Here the bracket ``[..]'' denotes
the antisymmetric combination, for example, $[su]=\frac{1}{\sqrt{2}}(su-us)$. Note, the flavor structure is common
for both tetraquark types, $|000\rangle$ and $|011\rangle$.}
\label{flavor structure}
\end{figure}

By construction, both tetraquarks have the same flavor structure, namely they form a nonet
which can be broken down to an octet and a singlet ($\bar{\bm{3}}_f \otimes \bm{3}_f = \bm{8}_f\oplus\bm{1}_f$).
Figure~\ref{flavor structure} shows a weight diagram for the flavor nonet with explicit individual quark
flavors~\footnote{One can easily construct this using a tensor notation. See Ref.~\cite{Oh:2004gz} for technical details.}.
Another common features of the two tetraquarks are that the spin($J$), parity($P$), charge conjugation($C$) are $J^{PC}=0^{++}$
and their isospins are restricted to $I=0,1/2,1$.

Our claim in Refs.~\cite{Kim:2016dfq,Kim:2017yur,Kim:2017yvd} is that the second tetraquark type must be considered
along with the first tetraquark
in the study of tetraquarks with $J^P=0^+$
because they mix strongly through the color-spin interaction~\cite{DeRujula:1975qlm,Keren07,Silve92,GR81},
\begin{eqnarray}
V_{CS} &=& v_0 \sum_{i < j}  \lambda_i \cdot \lambda_j \, \frac{J_i\cdot J_j}{m_i^{} m_j^{}}\label{CS}\ .
\end{eqnarray}
Here $\lambda_i$ denotes the Gell-Mann matrix for SU(3)$_c$, $J_i$ the spin, $m_i$ the constituent quark mass.
Because of this mixing, the hyperfine masses, which are the expectation values of the color-spin interaction with respect to $|000\rangle$, $|011\rangle$,
form a $2\times 2$ matrix~\footnote{To put this more precisely, one member belonging to
$|000\rangle$ and the other with the same flavor member belonging to $|011\rangle$
participate in making a $2\times 2$ matrix through the color-spin interaction.} for each isospin member.
The upshot of this mixing is that physical resonances can be identified by the eigenstates that diagonalize this $2\times 2$ matrix.
Since the quark mass and the color-electric terms are already diagonal almost proportional to the identity matrix in $|000\rangle$, $|011\rangle$,
the eigenstates of the hyperfine masses diagonalize the full Hamiltonian approximately.


\begin{table}
\centering
\begin{tabular}{c|l|c|c|c|c}  \hline\hline
    & $J^{PC}$ & $I$ &  Meson  & Mass(MeV) & $\Gamma$(MeV) \\
\hline
\multirow{2}{*}{Light} &
$0^{++}$ & $ 1 $ & $a_0 (980)$  & 980 &  50-100  \\
\multirow{2}{*}{nonet}&     $0^{+}$  & $1/2$ & $K_0^* (800)$ & 682 & 547  \\
& $0^{++}$ &   $ 0 $ & $f_0 (500)$  & 400-550 &  400-700   \\
& $0^{++}$ & $ 0 $ & $f_0 (980)$  & 990 &  10-100  \\

\hline
\multirow{2}{*}{Heavy} &
$0^{++}$ & $ 1 $ & $a_0 (1450)$ & 1474 &  265  \\
\multirow{2}{*}{nonet} &$0^{+}$ & $1/2$ & $K_0^* (1430)$ & 1425 & 270  \\
&$0^{++}$  & $ 0 $ & $f_0 (1370)$ & 1200-1500 &  200-500   \\
&$0^{++}$  & $ 0 $ & $f_0 (1500)$ & 1505 &  109   \\


\hline\hline
\end{tabular}
\caption{Tetraquark candidates are listed here for the light and heavy nonets collected from PDG~\cite{PDG16}.
Note that the experimental masses are different from the numbers appearing in the meson nomenclatures.}
\label{candidates}
\end{table}

This tetraquark mixing framework can be represented collectively by the formulas,
\begin{eqnarray}
|\text{Heavy~nonet} \rangle &=& -\alpha | 000 \rangle + \beta |011 \rangle \label{heavy}\ ,\\
|\text{Light~nonet} \rangle~ &=&\beta | 000 \rangle + \alpha |011 \rangle \label{light}\ ,
\end{eqnarray}
where the eigenstates, $|\text{Heavy~nonet} \rangle$ and $|\text{Light~nonet} \rangle$, are identified by the two nonets in PDG.
For the light nonet, we take the lowest-lying resonances in $J^P=0^+$, $a_0 (980)$, $K_0^* (800)$, $f_0 (500)$, $f_0 (980)$.
For the heavy nonet,
we take next higher resonances in $J^P=0^+$, $a_0 (1450)$, $K_0^* (1430)$, $f_0 (1370)$, $f_0 (1500)$.
These two nonets
are separated by huge mass gaps, more than 500 MeV or so, and they are also well separated from the rest higher resonances in $J^P=0^+$.
Their quantum numbers, experimental masses and widths are listed in Table~\ref{candidates}.
Table~\ref{parameters} presents our results for the mixing parameters, $\alpha$, $\beta$, and the hyperfine masses
calculated in the $|\text{Heavy~nonet} \rangle$, $|\text{Light~nonet} \rangle$ bases~\cite{Kim:2016dfq,Kim:2017yvd}.

\begin{table}[t]
\centering
\begin{tabular}{c|c||c|c||c|c}\hline\hline
  Light nonet & $\langle V_{CS} \rangle$ & Heavy nonet & $\langle V_{CS} \rangle$  & $\alpha$ & $\beta$ \\
\hline
 $a_0(980)$   & $-488.5$ &$a_0(1450)$   & $-16.8$ & 0.8167     & 0.5770    \\
 $K_0^*(800)$ & $-592.7$ &$K_0^*(1430)$ & $-26.9$ & 0.8130     & 0.5822    \\
 $f_0(500)$   & $-667.5$ &$f_0(1370)$   & $-29.2$ & 0.8136     & 0.5814    \\
 $f_0(980)$   & $-535.1$ &$f_0(1500)$   & $-20.1$ & 0.8157     & 0.5784    \\
\hline\hline
\end{tabular}
\caption{Here are the hyperfine masses and the mixing parameters, $\alpha, \beta$, associated with the configuration
mixing [Eqs.~(\ref{heavy}),(\ref{light})] collected from Refs.~\cite{Kim:2016dfq, Kim:2017yvd}.   For the isoscalar cases, we include
the flavor fixing according to ``Realistic case with fitting (RCF)''. See Ref.~\cite{Kim:2017yvd} for details. Note that
the hyperfine mass ordering, $\langle V_{CS} \rangle_{I=1} > \langle V_{CS} \rangle_{I=1/2}> \langle V_{CS} \rangle_{I=0} (\bm{8}_f)$, is the same
with the mass ordering in both nonets.  Here $\langle V_{CS} \rangle_{I=0} (\bm{8}_f)$ denotes the hyperfine mass of $f_0(500)$ or $f_0 (1370)$ depending on the nonets.}
\label{parameters}
\end{table}

This tetraquark mixing framework leads to the interesting outcomes
which can support the tetraquark structure of the two nonets~\cite{Kim:2016dfq,Kim:2017yur,Kim:2017yvd}.
Here we list some of the main results as the following.

\begin{enumerate}

\item One surprising result is the inequality among the mixing parameters, $\alpha > \beta$.
This implies that the light nonet members,
$a_0 (980)$, $K_0^* (800)$, $f_0 (500)$, $f_0 (980)$, have more probability to stay in
the configuration $|011\rangle$ rather than in $|000\rangle$.
This result is originated from the fact that the second tetraquark, $|011\rangle$, is more compact than the first tetraquark, $|000\rangle$,
namely $\langle 011 | V_{CS} |011 \rangle < \langle 000 | V_{CS} |000 \rangle$.
This is very different from a common expectation that
the light nonet is dominated by the configuration, $|000\rangle$.
But this result is supported
by the similar calculations~\cite{Black:1998wt} where this mixing
was used only to explain the small masses of the light nonet without identifying the other states in the heavy nonet.

\item Secondly, there is a strong mixing between $|000\rangle$, $|011\rangle$ through the color-spin interaction.
The mixing term, $\langle 000 | V_{CS} |011 \rangle$, is found to be very large, which, under the diagonalization of the
hyperfine mass matrix, separates
physical hyperfine masses by about 500 MeV or more between the two nonets. (see Table~\ref{parameters}.)
This is qualitatively consistent with the huge mass gaps, more than 500 MeV or so, existing between the two nonets in PDG.

\item A more direct outcome that can be tested in experiments is a peculiar characteristics in their decay modes
entirely from the tetraquark mixing framework.
When the two tetraquarks, $|000\rangle$, $|011\rangle$, decay into two pseudoscalar mesons through fall-apart mechanism,
our mixing framework predicts that the relative coupling
strengths are enhanced for the light nonet while they are suppressed for the heavy nonet.
In fact, this prediction is tested very well in the isovector channel, $a_0(980)$, $a_0(1450)$~\cite{Kim:2017yur}
in comparison with their experimental partial widths.
\end{enumerate}

\section{Further tetraquark signatures for the two nonets}
\label{sec:candidates}

Maybe one possible objection to the tetraquark mixing framework is the assumption that
the heavy nonet is the tetraquarks with flavor nonet similarly as the light nonet.
To solidify this assumption, it is necessary to collect all the tetraquark signatures.
The quantum numbers of the two nonets, $J^{PC}=0^{++}$, $I=0,1/2,1$, can be one of the signatures because they coincide
with those of the tetraquarks $|000\rangle$, $|011\rangle$.
But these quantum numbers can be generated also from other pictures like a two-quark picture with orbital angular
momentum $\ell=1$ so they are not the signatures exclusively for the tetraquarks.  More concrete signatures are
the items 2,3 above which say that the mass splittings and the fall-apart modes of the nonet members are
consistent with the tetraquark mixing framework.
In this section, we study additional tetraquark signatures for the two nonets.
The first part presents that the two nonets really form the flavor nonets with the mass ordering
consistent with the tetraquarks.  The second part is to show that our tetraquark mixing framework
can explain partially the marginal mass ordering seen in the heavy nonet.
In addition, we discuss the mixing parameters from the tetraquark mixing framework as additional
signature for the tetraquark.

\subsection{Two nonets as tetraquarks with the flavor nonet}
\label{sec:mass}

Let us begin by a discussion that the two nonets actually form the flavor nonets of SU(3)$_f$.
Apart from the quantum numbers, $J^{PC}=0^{++}$, $I=0,1/2,1$,
a more direct sign to support the flavor nonet can be seen from the Gell-Mann--Okubo(GMO) mass relation.
Using the experimental masses provided in Table~\ref{candidates}, we find that
the light octet satisfies the GMO relation,
$M^2[a_0(980)]+3M^2[f_0(500)]\approx 4M^2[K_0^*(800)]$, within 14
percent~\footnote{For the $f_0(500)$ mass, we take the central value from the mass range, $400-550$ MeV, given in Table~\ref{candidates}.}.
In addition,
$f_0(980)$ can be taken as another isoscalar belonging to the nonet simply because it is heavier than the octet counterpart, $f_0(500)$.
The heavy octet also satisfies the GMO relation, $M^2[a_0(1450)]+3M^2[f_0(1370)]\approx 4M^2[K_0^*(1430)]$, within 6
percent~\footnote{Again,
for the $f_0(1370)$ mass, we take the central value from its experimental mass range, $1200-1500$ MeV.}.
The other isoscalar $f_0(1500)$ is taken as a nonet member because it is heavier than $f_0(1370)$.
Therefore, the GMO relation provides a supporting evidence in taking the two nonets in PDG as the flavor nonets of SU(3)$_f$.
Our selection for $f_0(1370)$, $f_0(1500)$ may require further clarification because alternative suggestion
can be found in Ref.~\cite{Maiani:2006rq} where $f_0(1500)$ is a light glueball mixed with tetraquarks $f_0(1370)$, $f_0(1710)$.
But the lattice calculation~\cite{Sexton:1995kd} suggests that the scalar glueball is in good agreement with the observed
properties of $f_0(1710)$.

The GMO relation only shows that the two nonets form the flavor nonets but it
does not determine whether the flavor nonets belong to a two-quark system or a four-quark system.
More important characteristics in
identifying the two nonets as the tetraquark nonets is the mass ordering among the nonet members.
Here we reexamine this mass ordering in detail by considering the quark-mass contribution to the resonance masses.

Using the flavor wave functions given in Figure~\ref{flavor structure} while
assuming that $m_u=m_d\ne m_s$, it is straightforward to evaluate the quark mass contribution, $\sum m_{q}$, to
the mass of each isospin member as
\begin{eqnarray}
&&\sum m_{q}^{(I=1)}= 2m_s+2m_u \ ,\\
&&\sum m_{q}^{(I=1/2)}= m_s+3m_u \ ,\\
&&\sum m_{q}^{(I=0)} (\bm{8}_f)=\frac{2}{3}(m_s + 5m_u) \ .
\end{eqnarray}
In the last equation, $\sum m_{q}^{(I=0)} (\bm{8}_f)$ denotes the quark mass contribution to
the isoscalar resonance belonging to the flavor octet.
Now, by imposing $m_s > m_u$, one can establish the mass ordering among the quark mass contributions,
\begin{eqnarray}
\sum m_{q}^{(I=1)} ~> \sum m_{q}^{(I=1/2)}
 ~> \sum m_{q}^{(I=0)}(\bm{8}_f)\ .
\end{eqnarray}
This ordering should be maintained for the octet masses so that
our tetraquarks have the mass ordering
\begin{eqnarray}
M_{I=1} > M_{I=1/2}
> M_{I=0} (\bm{8}_f)\label{mass ordering}\ ,
\end{eqnarray}
among the octet members with definite isospins. This mass ordering, commonly known as ``inverted spectrum'', is a unique characteristics
of the tetraquarks clearly distinguished from a two-quark picture ($q\bar{q}$) which generates
the opposite ordering, $M_{I=1} < M_{I=1/2} < M_{I=0} (\bm{8}_f)$,
like the mass ordering seen in the pseudoscalar resonances, $m_\pi < m_K < m_\eta$.

The other isoscalar member belonging to the flavor singlet has not been listed in the ordering
Eq.~(\ref{mass ordering}) because of the following reasons.
Its mass $M_{I=0}(\bm{1}_f)$ lies between $M_{I=1}$ and $M_{I=1/2}$ if the ordering is governed by the quark mass term.
But the isoscalar masses can be further modified by the flavor mixing.
The flavor mixing, as it separates strange quarks from up and down quarks in the wave functions,
raises the flavor singlet mass, $M_{I=0} (\bm{1}_f)$, while pushes down
the flavor octet mass, $M_{I=0} (\bm{8}_f)$.
However, even under the flavor mixing, the ordering above, Eq.~(\ref{mass ordering}), is still
maintained whereas the position of $M_{I=0}(\bm{1}_f)$ in the ordering is slightly obscured.

For the light nonet, this mass ordering is clearly exhibited through
$M[a_0(980)] > M[K_0^*(800)] > M[f_0(500)]$ as one can see from Table~\ref{candidates}.
The other isoscalar $f_0(980)$ can be taken as a nonet member because it is heavier than $f_0(500)$, $K_0^*(800)$.
In fact, it is well known that the mass ordering
provides a clear ground in identifying the light nonet as tetraquarks~\cite{Jaffe77a,Jaffe77b}.
One may argue that the light nonet masses are rather small to be a four-quark state.
But, as we showed in Table~\ref{parameters}, the hyperfine masses
for the light nonet calculated from the tetraquark mixing framework
are huge negative numbers in the range $-670 \sim -490$ MeV, so that they can
qualitatively explain the smallness of the light nonet masses.

This mass ordering, Eq.~(\ref{mass ordering}), is also maintained in the heavy nonet.
The isovector member in the heavy nonet, $a_0(1450)$, is slightly heavier
than the isodoublet member, $K_0^*(1430)$, only by 50 MeV.
$K_0^*(1430)$ is heavier than the isoscalar
$f_0(1370)$ if its central value is taken from the experimentally known mass range $1200-1500$ MeV. (see Table~\ref{candidates}.)
So even though the mass ordering is marginal, one can assume the heavy nonet as tetraquarks.
Indeed, our tetraquark mixing framework in Refs.~\cite{Kim:2016dfq,Kim:2017yur,Kim:2017yvd},
where the heavy nonet is assumed to be tetraquarks,
provides nice phenomenological
agreements in terms of mass splitting and decay strengths as summarized in Sec.~\ref{sec:mixing framework}.
However, because of the marginal mass ordering, the heavy nonet may require further supports in treating its members as tetraquarks.

\subsection{Hyperfine mass ordering}
\label{sec:hyperfine ordering}

A related issue to the mass ordering is the ordering among the hyperfine masses
which can support our identification for the heavy nonet.
As one can see in Table~\ref{parameters}, our tetraquark mixing framework generates the hyperfine masses
which are ordered as $\langle V_{CS} \rangle_{I=1} > \langle V_{CS} \rangle_{I=1/2}> \langle V_{CS} \rangle_{I=0} (\bm{8}_f)$.
This ordering holds for the two nonets.
Note that this hyperfine mass ordering is the same as the mass ordering, Eq.~(\ref{mass ordering}).
This means that the mass ordering among the octet members is generated not only by the quark masses but also by the hyperfine masses.

But the magnitude of the spitting among the isospin members is quite different depending on the light and heavy nonets.
Their splittings in MeV unit obtained from Table~\ref{parameters} are
\begin{eqnarray}
\begin{array}{l|c|c}
     &   \text{Light nonet} & \text{Heavy nonet}  \\
\hline 
\langle V_{CS} \rangle_{I=1} - \langle V_{CS} \rangle_{I=1/2}  & 104.2    &  10.1   \\
\langle V_{CS} \rangle_{I=1/2} - \langle V_{CS} \rangle_{I=0} (\bm{8}_f) & 74.8     & 2.3
\end{array}
\nonumber\ .
\end{eqnarray}
From this, we see that the hyperfine masses contribute
to the mass ordering by 104 MeV or 75 MeV for the light nonet but they contribute only by 10 MeV or 2 MeV for the heavy nonet.
That is, the hyperfine mass splitting among the heavy nonet members is substantially narrower than the splitting among the light nonet members
so subsequently the mass ordering in the heavy nonet must be narrower by 94 MeV or 73 MeV.
This narrowing down in the hyperfine mass splitting
can provide a partial explanation on the marginal
mass ordering seen in the heavy nonet. Since this is a direct consequence of the mixing
formulas, Eqs.~(\ref{heavy}),(\ref{light}),
this result could be another support for our tetraquark mixing framework
that has not been pointed out in our previous works~\cite{Kim:2016dfq,Kim:2017yur,Kim:2017yvd}.

But we admit that such a narrowing down in hyperfine masses is not enough to explain the marginal mass ordering fully.
For a full description, it may be necessary to include a two-quark component in the heavy nonet
and its mixing with the tetraquarks in order to compensate the additional gap by the opposite mass ordering
generated from the two-quark component.
But this program may require anomalous interactions with the flavor determinant as the two-quark and tetraquarks do not mix
under the color-spin or color-electric interactions.
Anyway, based on the successful aspects seen in the tetraquark mixing framework,
we expect that the two-quark contribution is small. Nevertheless, this constitutes interesting future works to do.

\subsection{Mixing parameters}
\label{sec:mixing parameters}

Another signature to support the flavor nonet from the tetraquark mixing framework can be seen
through the mixing parameters, $\alpha$, $\beta$
in Eqs.~(\ref{heavy}),(\ref{light}).
The mixing parameters, $\alpha$, $\beta$, are fixed by the diagonalization of the hyperfine mass matrix {\it in each isospin channel}.
Their values, as they are determined in each isospin channel independently, must depend on isospins in principle.
But as one can see in Table~\ref{parameters},
the sensitivity to isospins are very small and in fact they are approximately close to the common values
$\alpha\approx \sqrt{2/3}=0.8165$, $\beta\approx\sqrt{1/3}=0.5774$.
The common mixing parameters are interesting given the fact that the associated hyperfine masses
have noticeable dependence on isospins as one can see in Table~\ref{parameters}.

These common mixing parameters imply
that the members in each nonet, given by either Eq.~(\ref{heavy}) or Eq.~(\ref{light}), are related by a SU(3)$_f$ rotation.
For example, an isovector member can be obtained by a SU(3)$_f$ rotation from an isodoublet member because both
members are represented approximately by the same mixing parameters, $\alpha$, $\beta$.
In other words, the left-hand sides of Eqs.~(\ref{heavy}), (\ref{light}) form
nonets governed approximately by SU(3)$_f$ just like $|000\rangle$, $|011\rangle$ do.
This is quite consistent with the phenomenological fact that the two nonets in PDG form the flavor nonets
through the GMO relation or the mass ordering.
Therefore, our result from the common mixing parameters provide another signature to support the tetraquark mixing framework.

\section{Other pictures for the two nonets}
\label{sec:other}

One may try different pictures other than tetraquarks in explaining the two nonets.
A two-quark picture($q\bar{q}$) with $\ell=1$ is one simple scenario as this can
generate the same quantum numbers of the two nonets.
Alternative picture that is often discussed in the literature is
hadronic molecules of meson-meson bound states.
In this section, we discuss the other pictures how their predictions are
different from the tetraquark picture and what the possible problems are.
So the purpose of the discussion here is to give a more orientation toward a tetraquark description for the two nonets.

\subsection{Two-quark picture with $\ell=1$}
\label{sec:two quark}

It is true that a two-quark($q\bar{q}$) picture with $\ell=1$ can generate a nonet with the quantum numbers $J^P=0^+$.
But we want to address that this picture does not fit well to the two nonets in PDG.
First of all, the number of possible configurations with the $q\bar{q}$ ($\ell=1$) picture is restricted to one so that this picture cannot
make the two nonets in the $J^P=0^+$ channel.
More precisely, when the spin of $q\bar{q}$, which we denote by $S$, is combined with the orbital angular momentum $\ell=1$, this
picture generate various nonets belonging to the total angular momentum, $J=0,1,2$, with the following configurations,
\begin{eqnarray}
&&J=0 :  (S=1, \ell=1)\label{j0} ,\\
&&J=1 :  (S=0, \ell=1) , (S=1, \ell=1)\label{j1} ,\\
&&J=2 :  (S=1, \ell=1) \label{j2} .
\end{eqnarray}
In this two-quark picture, the $J=0$ nonet must correspond to the two nonets in PDG but, as shown in Eq.~(\ref{j0}),
we have only one configuration ($S=1$, $\ell=1$), obviously not enough to explain the two nonets listed in PDG.
So the two nonets in PDG cannot be explained by the two-quark picture.

Alternatively, one may adopt different pictures for the light and heavy nonets.  Specifically,
one can assume that the light nonet is pure tetraquarks while the heavy nonet is described by a two-quark system
with $\ell=1$.  This assumption is based on an observation that
the light nonet clearly exhibits a well-separated mass ordering consistent with the tetraquarks while the heavy nonet
has the marginal ordering so that its tetraquark structure may be slightly obscured.
So, one may try a $q\bar{q}$ ($\ell=1$) picture only for the heavy nonet.

\begin{figure}[t]
\centering
\epsfig{file=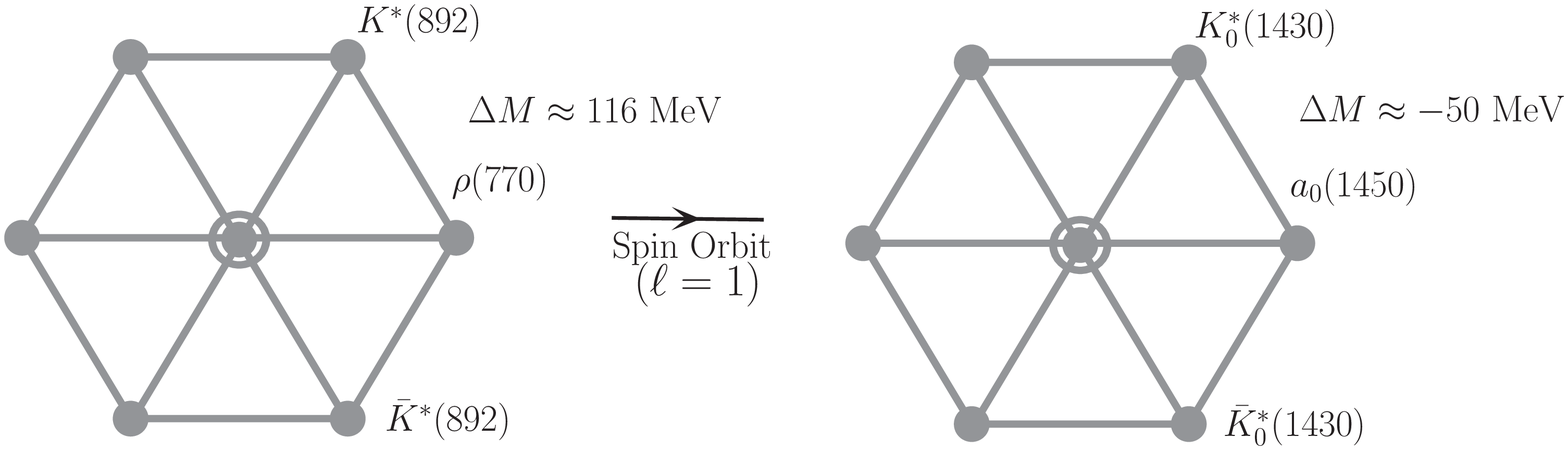, width=1.0\hsize}
\caption{A schematic picture to generate the heavy nonet
when its members are viewed as orbital excitations of the $q\bar{q}$ vector nonet. }
\label{two-quark picture}
\end{figure}

If the heavy nonet is viewed as a $q\bar{q}$ ($\ell=1$), this nonet must have the configuration ($S=1$, $\ell=1$) in making
the states with the total angular momentum, $J=0$.
This implies that the resonances in the heavy nonet can be regarded as the states orbitally excited from the
spin-1 vector mesons, $\rho, K^*,\omega, \phi$.  Thus,
the heavy nonet masses relative to the lowest-lying vector nonet must be generated by the spin-orbit interaction.
Problems with this two-quark picture can be seen from
the mass gaps between the $I=1/2, I=1$ members before and after turning on the spin-orbit interactions. As shown in Figure~\ref{two-quark picture},
its gap in the lowest-lying vector nonet is $M[K^*(892)]-M[\rho(770)] \approx 116$ MeV, with the isodoublet member being heavier.
Obviously, this mass ordering is driven by the strange quark being heavier than the up or down quark.
On the other hand, the corresponding mass gap in the heavy nonet, $M[K_0^* (1430)]-M[a_0 (1450)] \approx - 50$ MeV, exhibiting that
the isovector member is heavier.
To explain this, the spin-orbit interaction must contribute very differently to the $I=1,I=1/2$ members
so that it even flips the mass ordering established by the heavy strange quark mass.
We believe that this picture is not realistic for the heavy nonet.

Another approach is to introduce a mixing framework between the $q\bar{q}$ ($\ell=1$) and the tetraquark picture.
The marginal mass ordering
seen in the heavy nonet may have a room for a two-quark component to be included.
One way to accomplish this is to introduce a mixing between the $q\bar{q}$ ($\ell=1$) states with the tetraquark ($qq\bar{q}\bar{q}$) states
in generating the two nonets in PDG.  This is possible in principle because one can construct a SU(3)$_f$ invariant among the two nonets
represented by effective hadronic fields for
the $q\bar{q}$ and for the $qq\bar{q}\bar{q}$~\cite{Black:1999yz}.
But, as pointed out by Ref.~\cite{Maiani:2006rq}, the required mixing magnitude seems unnaturally large
given the fact that the very different configurations are involved.
In reality, the $q\bar{q}$ ($\ell=1$) states do not mix with the $qq\bar{q}\bar{q}$ states through the
color-spin and color-electric interactions which are, however, believed to be the major interactions among quarks inside hadrons.

\subsection{Meson-meson picture}
\label{sec:meson-meson}

Another picture to describe the two nonets in PDG is the meson-meson bound states
where two colorless mesons are bound by the color residual forces
just like a deuteron of proton-neutron system. This molecular states are expected to
have less binding than the tetraquarks whose binding is provided by direct
color forces between the two colorful objects, diquark and antidiquark.
Because of this, the meson-meson states, if they exist, must form shallow bound states.
This means that the meson-meson bound states can be characterized by their mass close to
the sum of their constituent meson masses~\cite{Guo:2017jvc,Wang:2017qbe}.
A typical example along this line is given by Ref.~\cite{Tornqvist:2004qy} where it is argued that
the $X(3872)$ mass, which is close to the $D^0 \bar{D}^{*0}$ threshold, may be an indication of its molecular nature.
One may adopt the similar picture for the light and heavy nonet members.
In particular, $f_0(980)$, $a_0(980)$ in the light nonet may be the shallow bound states
of $K\bar{K}$~\cite{Janssen:1994wn,Branz:2007xp,Branz:2008ha,Dudek:2016cru}
because their masses are in the proximity of $2M_K$. Also for the heavy nonet, one can build the
shallow bound states from two vector mesons whose total masses are not far from masses of the heavy nonet members.
But this molecular picture may not be applied to all the members in both nonets. For example,
$f_0 (500)$ in the light nonet is hard
to be considered as a shallow bound state of $\pi\pi$ due to the large mass difference.

In practice, the meson-meson picture tends to involve model dependence as its description relies
on some phenomenological pictures such as one pion exchange potential.
Binding mechanisms as well as the corresponding energies
may not be determined unanimously.  So this picture is difficult to be conclusive.
Another problem with the meson-meson picture is absence of the additional multiplets expected in the hadron spectrum.
The bound states from meson octet-meson octet form
various multiplets through the group multiplication,
$\bm{8}\otimes \bm{8}=\bm{27}\oplus \bm{10}\oplus \bm{\overline{10}}\oplus \bm{8}\oplus \bm{8}\oplus \bm{1}$.
The color residual forces that bind these mesons are of the same type because the constituent mesons are always
in the color-singlet state of $\sim \bar{q}^a q_a$. In other words, the color residual forces do not discriminate much
in forming all the possible multiplets.
So, if this meson-meson picture works, we expect the additional resonances with the higher isospins $I=2,3/2$
belonging to $\bm{27}$, $\bm{10}$ or $\bm{\overline{10}}$ in addition to
the two nonets with $I=0,1/2,1$.  But this expectation is not supported by the current PDG.
Even the states with $I=0,1/2,1$ need to be more numerous than what we currently have in PDG
as the higher multiplets can generate those
isospin states also. (see Table I in Ref.~\cite{Kim:2017yvd} for the $0^+$ resonances in PDG.)
One possible resolution to this discrepancy can be found from Ref.~\cite{Hyodo:2006yk,Hyodo:2006kg}
where the attraction in those exotic channels, calculated through Weinberg-Tomozawa term in the flavor SU(3)$_f$ symmetric limit,
is not strong enough to generate bound states.  So the meson-meson picture may have some room to accommodate the two nonets still.

Our tetraquark mixing model has one common feature with this meson-meson picture in that the tetraquark
wave functions also have two-meson channels which can fall apart into either two pseudoscalar or two vector mesons.
The difference is that the two-meson channels are only the part of the wave functions in the tetraquark picture while they are
the full component in the meson-meson picture.  In addition, the two-meson channels are close to on-shell in the meson-meson picture
while they can be extended to off-shell in the tetraquark picture.
More importantly,
the tetraquark mixing model predicts quite different coupling strengths for the two-pseudoscalar modes
depending on the light or heavy nonet~\cite{Kim:2017yur,Kim:2017yvd}.
This could be a unique consequence of the tetraquark mixing framework generated by Eqs.~(\ref{heavy}),(\ref{light}),
probably not present in the meson-meson picture as there is no similar mixing mechanism.
As we will examine below in Sec.~\ref{sec:fall-apart}, one can establish similar phenomena in the two vector couplings
with quite different characteristics.
Eventually, this type of studies can be used to determine a realistic picture for the two nonets, meson-meson picture
or tetraquark mixing picture.

\section{Fall-apart modes into two vector mesons}
\label{sec:fall-apart}

Our tetraquark mixing framework can be succinctly represented by Eqs.~(\ref{heavy}),(\ref{light}).
Due to the relative sign difference in the two equations, the two tetraquark types, $|000\rangle$, $|011\rangle$,
partially cancel in making the heavy nonet while they add up in making the light nonet.
This in fact leads to an interesting phenomenon in the fall-apart strengths into
two pseudoscalar mesons~\cite{Kim:2017yur}.  Specifically, the corresponding coupling strengths are suppressed for the heavy nonet while
they are enhanced for the light nonet. In this section, we look for another signature for
the tetraquark mixing framework from the fall-apart modes into the two vector mesons.

First, the appearance of two-meson states can be demonstrated easily by rewriting the tetraquark wave functions
with respect to the quark and antiquark bases.
To illustrate this, let us label the tetraquarks in the diquark-antidiquark picture as
$q_1 q_2 \bar{q}_3\bar{q}_4$ and rewrite them by recombining quarks and antiquarks into the $13$- and
$24$-pair~\footnote{The other recombination into the $14$- and $23$-pair gives the same fall-apart modes.}.
This recombination in color space is schematically represented by
\begin{eqnarray}
(q_1 q_2 \bar{q}_3\bar{q}_4)_{\bm{1}_c} \sim \left [\bm{8}_{c13}\otimes \bm{8}_{c24} \right ]_{\bm{1}_c}
+ \left [\bm{1}_{c13}\otimes \bm{1}_{c24} \right ]_{\bm{1}_c}\label{recomb} .
\end{eqnarray}
Here the notation such as $\bm{8}_{c13}$ denotes the $13$-pair of quark-antiquark in the color octet.
From this equation, we notice that the tetraquarks have two components
differed by the color configurations in the quark-antiquark bases. The first
component is composed by two pairs of quark-antiquark, both belonging to $\bm{8}_{c}$ in forming a color-singlet totally.
The second component contains the two pairs belonging to $\bm{1}_{c}$.

It is this second component that corresponds to two-meson modes which
can fall apart into two mesons if the decays are kinematically allowed.
Specifically the inner product of a two-meson state
with Eq.~(\ref{recomb}) picks out the same state from the second component with
the corresponding relative coupling determined by the color, spin, and flavor recombination factors.
Thus, to find the relative strengths of the fall-apart modes, one needs to calculate the numerical factors coming from
color, spin and flavor recombination separately in the second term of Eq.~(\ref{recomb}).
The purpose of this section is to provide technical details in calculating
possible modes and their relative strengths in two vector channels.

\subsection{Color and spin factors in the recombination}
\label{subsec:color-spin factors}

Let us start with the color factors in this recombination of the tetraquark wave functions.
To form two mesons in the final states,
the $13$- and $24$-pair in Eqs.~(\ref{color 1}), (\ref{color 2}) must be in a color-singlet state separately.
By isolating the color-singlet pieces, and substituting them into Eqs.~(\ref{color 1}), (\ref{color 2}),
one arrives at the following replacements
\begin{eqnarray}
&&\frac{1}{\sqrt{12}}  \varepsilon_{abd}^{} \ \varepsilon_{aef}
\Big ( q^b_1 q^d_2 \Big )
\Big ( \bar{q}^e_3 \bar{q}^f_4 \Big )\
\rightarrow  \frac{1}{\sqrt{3}} \bm{1}_{c13} \bm{1}_{c24}\label{color1_factor} ,\\
&&\frac{1}{\sqrt{96}} \Big( q^a_1 q^b_2+q^b_1 q^a_2 \Big )
\Big (\bar{q}^a_3 \bar{q}^b_4+\bar{q}^b_3 \bar{q}^a_4\Big )
\rightarrow  \sqrt{\frac{2}{3}} \bm{1}_{c13} \bm{1}_{c24} .
\label{color2_factor}
\end{eqnarray}
This shows that the recombination color factor is $\sqrt{1/3}$ from the first tetraquark type, $|000\rangle$,
and $\sqrt{2/3}$ from the second tetraquark type, $|011\rangle$. An additional sign expected from anticommuting fermion fields
does not affect our results because it changes only the overall sign for both Eqs.~(\ref{color1_factor}), (\ref{color2_factor}).
Our results in the fall-apart modes depend only on the relative signs.

One thing to point out is that the color recombination factors are approximately close to the
tetraquark mixing parameters, $\alpha\approx \sqrt{2/3}$, $\beta\approx \sqrt{1/3}$.  If these numbers are
inserted in Eq.~(\ref{heavy}), the fall-apart modes almost cancel eventually yielding zero coupling
strengths to the heavy nonet.  Thus, as far as the color factors are concerned, the fall-apart modes
from the heavy nonet vanish approximately.  However, this expectation does not occur when
the additional factors coming from the spin recombination below are included.

Next, we calculate the spin factors in this recombination.
The first tetraquark has the spin configuration, $|J J_{12} J_{34}\rangle = |000\rangle$.
The diquark, since its spin $J_{12}$ and the spin projection $M_{12}$ are zero, is in the state,
$|J_{12} M_{12}\rangle=|0,0\rangle_{12}$. And the antidiquark is
in $|J_{34} M_{34}\rangle=|0,0\rangle_{34}$. This means, the first tetraquark can be written as
\begin{eqnarray}
|000\rangle = |0,0\rangle_{12} |0,0\rangle_{34}\ ,
\end{eqnarray}
in terms of the diquark (antidiquark) spin and its projection.
Since $|0,0\rangle_{12}=\frac{1}{\sqrt{2}} [ \uparrow_1 \downarrow_2 -\downarrow_1 \uparrow_2 ]$,
$|0,0\rangle_{34}=\frac{1}{\sqrt{2}} [ \uparrow_3 \downarrow_4 -\downarrow_3 \uparrow_4 ]$,
one can rewrite $|000\rangle$ in terms of the individual quark spinors as
\begin{eqnarray}
|000\rangle = \frac{1}{2} [ \uparrow_1 \downarrow_2 -\downarrow_1 \uparrow_2 ]~[ \uparrow_3 \downarrow_4 -\downarrow_3 \uparrow_4 ]\label{000_a}\ .
\end{eqnarray}
The $13$-pairs in the right-hand side can be expressed by the definite spin states, $|J_{13} M_{13}\rangle$, as
\begin{eqnarray}
\uparrow_1 \uparrow_3 &=& |1,1\rangle_{13} \label{13_1}\ ,\\
\uparrow_1 \downarrow_3 &=& \frac{1}{\sqrt{2}} [|1,0\rangle_{13} + |0,0\rangle_{13}]\label{13_2}\ ,\\
\downarrow_1 \uparrow_3 &=& \frac{1}{\sqrt{2}} [|1,0\rangle_{13} - |0,0\rangle_{13}]\label{13_3}\ ,\\
\downarrow_1 \downarrow_3 &=& |1,-1\rangle_{13}\label{13_4} \ .
\end{eqnarray}
The $24$-pairs in Eq.~(\ref{000_a}) can be written similarly by its spin states,
$|1,1\rangle_{24}$, $|1,0\rangle_{24}$, $|1,-1\rangle_{24}$, $|0,0\rangle_{24}$.

Putting all these into Eq.~(\ref{000_a}), we obtain the final expression for $|000\rangle$ with respect to the spin states of the $13$-, $24$-pair
\begin{eqnarray}
|000\rangle &=&\frac{1}{2}\Big [ |1,1\rangle_{13} |1,-1\rangle_{24} - |1,0\rangle_{13} |1,0\rangle_{24}  \nonumber \\
            & &+ |0,0\rangle_{13} |0,0\rangle_{24} + |1,-1\rangle_{13} |1,1\rangle_{24} \Big ]\label{000_1}\ .
\end{eqnarray}
The subscripts, ``$13$'' and ``$24$'', denote quark-antiquark pairs so each term in the right-hand side corresponds to two-meson channels
with the designated spin states.
The component, $|0,0\rangle_{13} |0,0\rangle_{24}$, corresponds to two pseudoscalar mesons
and other three components,  $|1,1\rangle_{13} |1,-1\rangle_{24}$, $|1,0\rangle_{13} |1,0\rangle_{24}$, $|1,-1\rangle_{13} |1,1\rangle_{24}$,
correspond to two vector-mesons distinguished by the spin projections.  These modes can be measured experimentally if
the invariant masses of two mesons are less than the tetraquark masses.
By defining the two pseudoscalar and two vector parts as
\begin{eqnarray}
PP&=&|0,0\rangle_{13} |0,0\rangle_{24}\ ,\\
VV&=& \frac{1}{\sqrt{3}}\Big [ |1,1\rangle_{13} |1,-1\rangle_{24} - |1,0\rangle_{13} |1,0\rangle_{24}\nonumber\\
  & & + |1,-1\rangle_{13} |1,1\rangle_{24} \Big ] \ ,
\end{eqnarray}
one can rewrite Eq.~(\ref{000_1}) neatly as
\begin{eqnarray}
|000\rangle &=&\frac{1}{2}PP + \frac{\sqrt{3}}{2} VV\label{000}\ .
\end{eqnarray}

The second tetraquarks have the spin configuration, $|011\rangle$, containing
spin-1 diquark and spin-1 antidiquark with three possible projections.
The four-quark state, as its spin and projection are zero,
can be expressed by the spin states of diquark and antidiquark as
\begin{eqnarray}
|011\rangle &=& \frac{1}{\sqrt{3}} \Big \{ |1,1\rangle_{12} |1,-1\rangle_{34} - |1,0\rangle_{12} |1,0\rangle_{34} \nonumber \\
& &+ |1,-1\rangle_{12} |1,1\rangle_{34}] \Big \}\label{011_a}\ ,
\end{eqnarray}
including the relevant Clebsch-Gordan coefficients.
The rearrangement into the $13$-, $24$-pair can be done similarly as before by using Eqs.~(\ref{13_1}), (\ref{13_2}), (\ref{13_3}), (\ref{13_4}).
Skipping all the details, we simply write down the final expression,
\begin{eqnarray}
|011\rangle &=&\frac{1}{2\sqrt{3}}\Big \{ -|1,1\rangle_{13} |1,-1\rangle_{24} + |1,0\rangle_{13} |1,0\rangle_{24} \nonumber \\
& & + 3|0,0\rangle_{13} |0,0\rangle_{24} - |1,-1\rangle_{13} |1,1\rangle_{24} \Big \}\nonumber \\
&=&\frac{\sqrt{3}}{2}PP - \frac{1}{2} VV
\label{011}\ .
\end{eqnarray}
Similarly as Eq.~(\ref{000}), this 2nd tetraquark can couple to two pseudoscalar mesons and two vector mesons also.

Comparing Eqs.~(\ref{000}),(\ref{011}), one can see that the two-pseudoscalar mode, $PP$,
has the same sign in both equations but with different numerical factors.
Substituting Eqs.~(\ref{000}),(\ref{011}) into Eq.~(\ref{heavy}),
we see in the heavy nonet that the cancelation still occurs for the two-pseudoscalar modes but partially.
For the light nonet, by inserting Eqs.~(\ref{000}),(\ref{011}) into Eq.~(\ref{light}), we see that
the two-pseudoscalar modes add up each other.
This difference eventually makes the associate couplings
to be suppressed for the heavy nonet and enhanced for the light nonet~\cite{Kim:2017yur,Kim:2017yvd}.

On the other hand, the two-vector mode, $VV$, has the opposite sign in Eqs.~(\ref{000}),(\ref{011}).
In this case, we find analogous results but with the {\it opposite trend}.
Namely, the two-vector modes add up in the heavy nonet enhancing their associate couplings
while they cancel each other in the light nonet suppressing their couplings.
So this characteristics in the two-vector modes is clearly different from the two-pseudoscalar modes.

By combining the color factors in Eqs.~(\ref{color1_factor}), (\ref{color2_factor}),
with the spin factors in Eqs.~(\ref{000}), (\ref{011}), we obtain the corresponding factors for the two-vector modes
\begin{eqnarray}
|000\rangle &\rightarrow& \frac{1}{\sqrt{3}} \times \frac{\sqrt{3}}{2} = \frac{1}{2}\label{number1}\ ,\\
|011\rangle &\rightarrow& \sqrt{\frac{2}{3}} \times \left(-\frac{1}{2}\right) = -\frac{1}{\sqrt{6}}\label{number2}\ ,
\end{eqnarray}
from the two tetraquark types.
Inserting these into Eqs.~(\ref{heavy}),(\ref{light}), we finally obtain the necessary factors
coming from the color and spin recombination for the heavy and light nonet respectively as
\begin{eqnarray}
|\text{Heavy nonet}\rangle &\rightarrow& -\frac{\alpha}{2} - \frac{\beta}{\sqrt{6}}\label{color-spin1}\ ,\\
|\text{Light nonet}\rangle &\rightarrow& \frac{\beta}{2} - \frac{\alpha}{\sqrt{6}}\label{color-spin2}\ .
\end{eqnarray}
As one can see from the relative signs, the two terms add up in Eq.~(\ref{color-spin1}) while the corresponding terms
cancel partially in Eq.~(\ref{color-spin2}).
This form of the color and spin factors is common for all the members in each nonet even
though $\alpha$, $\beta$ have slight isospin dependence as shown in Table~\ref{parameters}.
Therefore, as advertised, the two-vector modes are enhanced for the heavy nonet while they are suppressed for the light nonet.

\subsection{Flavor factors in the recombination}

To find concrete two-meson channels,
we now move to the flavor recombination in terms of the $13$-, $24$-pair.
Our tetraquarks have the flavor structures given in Figure~\ref{flavor structure}
and they are the same for $|000\rangle$, $|011\rangle$ by construction.  In order to find the fall-apart modes into the two vector mesons,
we start by writing the $q\bar{q}$ representation for the vector mesons,
\begin{eqnarray}
&&u\bar{s}=K^{*+},\ d\bar{s}=K^{*0},\ s\bar{u}=\bar{K}^{*-},\ s\bar{d}=\bar{K}^{*0}\nonumber\ ,\\
&&s\bar{s}=\phi,\ u\bar{u}=\frac{1}{\sqrt{2}}(\omega+\rho^0),\ d\bar{d}=\frac{1}{\sqrt{2}}(\omega-\rho^0)\nonumber\ ,\\
&&u\bar{d}=\rho^+,\ d\bar{u}=\rho^-\label{overlap}\ .
\end{eqnarray}
Here the flavor mixing is assumed to occur ideally according to Okubo-Zweig-Iizuka(OZI) rule
so that the strange quark is completely decoupled from the up and down quarks
among the two isoscalar members, $\phi$, $\omega$.

Now for the isovector members in our tetraquarks, we have $a_0^+(980)$ in the light nonet and $a_0^+(1450)$
in the heavy nonet~\footnote{We choose the positive-charged state here but
our result should be the same for the other isospin members as they are simply related by isospin rotations.}.
Both have the common flavor structure $[su][\bar{d}\bar{s}]$.
In the recombination into the $13$-, $24$-pair,
it becomes $(s\bar{d})(u\bar{s})-(s\bar{s})(u\bar{d})$.
Using the identifications in Eq.~(\ref{overlap}), we find the following two-vector modes common for $a_0^+(980)$, $a_0^+(1450)$,
\begin{eqnarray}
\bar{K}^{*0}K^{*+}- \phi \rho^+ \label{isovector_f}\ .
\end{eqnarray}

For the isodoublet members, we have $K_0^{*+}(800)$ in the light nonet and $K_0^{*+}(1430)$ in the heavy nonet.
Their flavor structure, which is common for both, is $[ud][\bar{d}\bar{s}]$.
If it is rearranged into the quark-antiquark pairs, we find $(u\bar{d})(d\bar{s})-(u\bar{s})(d\bar{d})$.
Again the representation in Eq.~(\ref{overlap}) leads to the two-vector modes common for $K_0^{*+}(800)$, $K_0^{*+}(1430)$,
\begin{eqnarray}
\rho^+ K^{*0} -\frac{1}{\sqrt{2}} K^{*+} \left [ \omega - \rho^0\right ] \label{isodoublet_f}\ .
\end{eqnarray}

For the isoscalar members, we have two sets of resonances.
One set is $f_0(500)$, $f_0 (1370)$, which are close to the flavor octet members belonging separately to the two nonets,
and the other set is $f_0 (980)$, $f_0 (1500)$, which are close to the flavor singlet members.
The two resonances in each set have the same flavor structure.
The reason why we call these resonances as ``close to the octet or singlet'' is because they are subject to the flavor mixing
according to the OZI rule generalized to the four-quark
system~\footnote{This type of generalization of OZI rule has been applied also to pentaquarks~\cite{Lee:2004bsa}.}.
In other words, they are not definite members of SU(3)$_f$.
The flavor mixing occurs among the light nonet members, $f_0(500)$, $f_0 (980)$, and similarly
among the heavy nonet members, $f_0 (1370)$, $f_0 (1500)$.

In this generalization of the flavor mixing, the isoscalar resonances can be represented by linear combinations
of the ideal mixing states, $|L\rangle$ $|H\rangle$, defined by
\begin{eqnarray}
|L\rangle &=&  [ud][\bar{u}\bar{d}]  \label{low}\ ,\\
|H\rangle &=& \frac{1}{\sqrt{2}} \{ [ds][\bar{d}\bar{s}] + [su][\bar{s}\bar{u}] \} \label{high}\ .
\end{eqnarray}
Namely, the flavor structure of
the first set of resonances, $f_0(500), f_0 (1370)$, is given by the combination
\begin{eqnarray}
a |L\rangle + b | H\rangle\label{lower}\ ,
\end{eqnarray}
while the second set of resonances, $f_0(980), f_0 (1500)$, have the flavor structure written by
\begin{eqnarray}
-b |L\rangle + a | H\rangle\label{higher}\ .
\end{eqnarray}
The flavor mixing parameters, $a,b$, can be fixed depending on how we implement the flavor mixing~\cite{Kim:2017yvd}.
When $a=\sqrt{2/3}$, $b=-\sqrt{1/3}$, we get the SU(3)$_f$ symmetric case where there is no flavor mixing.
When $a=1$, $b=0$, we get the ideal mixing case, i.e., $|L\rangle$, $| H\rangle$. In this case,
the four-quark states containing strange quarks are completely separated from those states
composed only by up and down quarks.
The other case is the realistic case with fitting (RCF)~\cite{Kim:2017yvd} where we fit the flavor mixing parameters by
tuning them to reproduce the mass splitting between $f_0 (980)$, $f_0 (1500)$ equivalent to their hyperfine mass splitting.
The flavor mixing parameters in RCF have been determined to be
\begin{eqnarray}
a=0.8908, ~~b=-0.4543\label{RCF par}\ .
\end{eqnarray}

Among the three cases,
the physical isoscalar states in each nonet must be the ones that diagonalize the full Hamiltonian.
Within our framework using the color-spin interaction, $V_{CS}$,
the two isoscalars in each nonet can be mixed and further diagonalization is necessary in order to generate the physical states.
In fact, one can demonstrate that the ideal mixing states, $|L\rangle$, $|H\rangle$, are
the eigenstates that diagonalize the color-spin interaction as $\langle L| V_{CS}| H \rangle = 0$.
However, neither the SU(3)$_f$ symmetric case nor the ideal mixing case represent the real situation for the isoscalars
in the two nonets.  In reality, there could be additional interactions like the anomalous term which is normally responsible for
the mass splitting between $\eta$ and $\eta^\prime$.  Such an interaction may exist also in our tetraquark system.
But, according to our analysis for the decays into two pseudoscalar
mesons~\cite{Kim:2017yvd}, some modes in the ideal mixing case or in
the SU(3)$_f$ symmetric case are found to be inconsistent with
the experimental modes.  Moreover, the anomalous interaction introduces further ambiguity in the model as additional
parameters entailed in this interaction may not be estimated reliably.
Instead, RCF includes such an interaction indirectly as the flavor mixing parameters
here are fitted to the physical mass difference of two
isoscalar resonances.
The RCF result seems to give a more consistent description of the two-pseudoscalar fall-apart modes in comparison with
the PDG data.  So in our fall-apart modes into two vector mesons, we will be considering this realistic case only.

Now, the fall-apart modes for the isoscalar resonances can be constructed from those for $|L\rangle$, $|H\rangle$
through Eqs.~(\ref{lower}), (\ref{higher}) with the mixing parameters given in Eq.~(\ref{RCF par}).
In fact, the fall-apart modes for $|L\rangle$, $|H\rangle$ are determined straightforwardly as
\begin{eqnarray}
|L\rangle &\Rightarrow& \frac{1}{2} \left [ \omega \omega - \bm{\rho}\cdot\bm{\rho} \right ]\label{modes_L}\ ,\\
|H\rangle &\Rightarrow& \omega \phi - \frac{1}{\sqrt{2}}\overline{K}^{*} K^*\label{modes_H}\ ,
\end{eqnarray}
with the short-hand notations for the isovector and isodoublet resonances defined by
\begin{eqnarray}
&&\bm{\rho}\cdot\bm{\rho}=\rho^0\rho^0 + \rho^+\rho^- + \rho^-\rho^+\nonumber\ ,\\
&&\overline{K}^{*} K^*=K^{*-} K^{*+}+\bar{K}^{*0} K^{*0}\nonumber \ .
\end{eqnarray}
Putting these modes into Eq.~(\ref{lower}), we find
the two-vector modes for $f_0(500)$, $f_0 (1370)$ as
\begin{eqnarray}
\frac{a}{2} \omega\omega + b\omega\phi
-\frac{a}{2} \bm{\rho}\cdot\bm{\rho} -\frac{b}{\sqrt{2}}\overline{K}^{*} K^*\label{isoscalar1_f}\ .
\end{eqnarray}
Note that, due to the isospin factors, the $\rho^+\rho^-$ coupling is twice of the $\rho^0\rho^0$ coupling and the $K^{*-} K^{*+}$ coupling is the
same with the $\bar{K}^{*0} K^{*0}$ coupling.

The two-vector modes from the other members, $f_0(980)$ and $f_0 (1500)$,
are obtained from Eq.~(\ref{isoscalar1_f}) simply by replacing $a\rightarrow -b$, $b\rightarrow a$,
\begin{eqnarray}
-\frac{b}{2} \omega\omega + a\omega\phi +\frac{b}{2} \bm{\rho}\cdot\bm{\rho} -\frac{a}{\sqrt{2}}\overline{K}^{*} K^*\label{isoscalar2_f}\ ,
\end{eqnarray}
because their flavor structure is given by Eq.~(\ref{higher}) that can be obtained from Eq.~(\ref{lower}) with the same replacements.

\subsection{Total strengths of the fall-apart modes}

Relative strengths of the fall-apart modes into two vector mesons can be obtained by combining the
spin and color recombination factor, Eq.~(\ref{color-spin1}) for the heavy nonet
and Eq.~(\ref{color-spin2}) for the light nonet, with the flavor recombination factor, Eq.~(\ref{isovector_f}) for the isovectors,
Eq.~(\ref{isodoublet_f}) for the isodoublets,
and Eqs.~(\ref{isoscalar1_f}), ~(\ref{isoscalar2_f}) for the isoscalar resonances. They are given by
\begin{widetext}
\begin{eqnarray}
a_0^+ (980) &:~&
\Big\{\bar{K}^{*0}K^{*+}- \phi \rho^+ \Big\}
\times \left ( \frac{\beta}{2} - \frac{\alpha}{\sqrt{6}} \right ) \label{a0(980)mode}\ ,\\
a_0^+ (1450) &:~&
\Big\{\bar{K}^{*0}K^{*+}- \phi \rho^+ \Big\}
\times \left (-\frac{\alpha}{2} - \frac{\beta}{\sqrt{6}} \right )\label{a0(1450)mode}\ ,\\
K^{*+}_0 (800) &:~&
\left \{\rho^+ K^{*0} -\frac{1}{\sqrt{2}} K^{*+} ( \omega - \rho^0 ) \right\}
\times \left ( \frac{\beta}{2} - \frac{\alpha}{\sqrt{6}} \right )\label{K0(800)mode}\ ,\\
K^{*+}_0 (1430) &:~&
\left\{\rho^+ K^{*0} -\frac{1}{\sqrt{2}} K^{*+} ( \omega - \rho^0 ) \right\}
\times \left (-\frac{\alpha}{2} - \frac{\beta}{\sqrt{6}} \right )\label{K0(1430)mode}\ ,\\
f_0(500) &:~&
\Big\{\frac{a}{2} \omega\omega + b\omega\phi
-\frac{a}{2} \bm{\rho}\cdot\bm{\rho} -\frac{b}{\sqrt{2}}\overline{K}^{*} K^*
\Big\}\times \left ( \frac{\beta}{2} - \frac{\alpha}{\sqrt{6}} \right )\label{500mode}\ ,\\
f_0(1370) &:~&
\Big\{\frac{a}{2} \omega\omega + b\omega\phi
-\frac{a}{2} \bm{\rho}\cdot\bm{\rho} -\frac{b}{\sqrt{2}}\overline{K}^{*} K^*
\Big\}\times \left (-\frac{\alpha}{2} - \frac{\beta}{\sqrt{6}} \right )\label{1370mode}\ ,\\
f_0(980) &:~&
\Big\{-\frac{b}{2} \omega\omega + a\omega\phi +\frac{b}{2} \bm{\rho}\cdot\bm{\rho} -\frac{a}{\sqrt{2}}\overline{K}^{*} K^*
\Big\}\times \left ( \frac{\beta}{2} - \frac{\alpha}{\sqrt{6}} \right )\label{980mode}\ ,\\
f_0(1500) &:~&
\Big\{-\frac{b}{2} \omega\omega + a\omega\phi +\frac{b}{2} \bm{\rho}\cdot\bm{\rho} -\frac{a}{\sqrt{2}}\overline{K}^{*} K^*
\Big\} \times \left (-\frac{\alpha}{2} - \frac{\beta}{\sqrt{6}} \right )\label{1500mode}\ .
\end{eqnarray}
\end{widetext}
For each isospin channel, the flavor recombination factor is the same for both nonets because the two nonets have the same flavor structure.
So the fall-apart modes from the heavy and light nonets differ by color and spin recombination factors only.

The tetraquark mixing parameters, $\alpha, \beta$, depend on isospins and their numerical values
can be found in Table~\ref{parameters}.  The flavor mixing parameters, $a,b$, are given in Eq.~(\ref{RCF par}).
Using them, we present in Table~\ref{couplings} the numerical values for the fall-apart coupling strengths
from all the resonances in the two nonets. It should be remembered that these are the relative strengths and the actual
strengths are supposed to be multiplied by an unknown overall constant.
Table~\ref{couplings} clearly shows that the relative
strengths are enhanced for the heavy nonet while they are suppressed for the light nonet.
The ratios of the two strengths, one from the light nonet and the other from the heavy nonet,
become huge numbers around $\sim 15$ and they do not suffer from the unknown overall constant.
One can also see that all the fall-apart modes in each isospin channel yield the same ratio.
The ratio is 14.33 for $I=1$,
15.78 for $I=1/2$, 15.54 for $I=0 (\sim \bm{8}_f)$, and 14.70 for $I=0 (\sim \bm{1}_f)$.
This is a direct consequence of the fact that the two nonets have the same flavor structure.
Namely, for each fall-apart mode, the heavy and light nonets have the same flavor recombination factor
which is canceled away from the ratios.
So the ratios are fixed purely by the spin and color recombination factors
through Eq.~(\ref{color-spin1}) over Eq.~(\ref{color-spin2}).


\begin{table}
\centering
\begin{tabular}{c|c|c|c|c}  \hline\hline
 & mode   & $a_0^+(980)$  &  $a_0^+(1450)$ & ratio \\
\cline{2-5} \\[-3.0mm]
$I=1$ & $\bar{K}^{*0}K^{*+}$ &  $ -0.0449 $   & $-0.6439$ &  14.33 \\
 & $\phi \rho^+$ &  $ 0.0449 $  & $0.6439$ &  \\
\hline
 &  mode & $K_0^{*+}(800)$ &  $K_0^{*+}(1430)$ & ratio \\
\cline{2-5} \\[-3.0mm]
& $\rho^+ K^{*0}$&  $ -0.0408 $ & $-0.6442$  & \\
$I=1/2$ & $\rho^0 K^{*+}$&  $ -0.0289 $ & $-0.4555$ & 15.78 \\
 &$\omega K^{*+} $ & $0.0289$ & $0.4555 $ & \\
\hline
 & mode   & $f_0(500)$  &  $f_0(1370)$ & ratio \\
\cline{2-5} \\[-3.0mm]
&$\rho^0\rho^0$ &   $ 0.0185 $ & $0.2869$ &  \\
$I=0$ &$\bar{K}^{*0} K^{*0}$ &   $ -0.0133 $ & $-0.2069$ & 15.54 \\
($\sim \bm{8}_f$) &$\phi \omega$  &   $ 0.0188 $  & $0.2927$ &   \\
&$\omega \omega$ &  $-0.0185$  & $-0.2869 $ & \\
\hline
 & mode & $f_0(980)$ &  $f_0(1500)$ & ratio \\
\cline{2-5} \\[-3.0mm]
 &$\rho^0\rho^0$  &  $ 0.0100 $  & $0.1463$ &  \\
$I=0$ &$\bar{K}^{*0} K^{*0}$  &   $ 0.0276 $  & $0.4057$ & 14.70 \\
($\sim \bm{1}_f$) &$\phi \omega$ &   $ -0.0390 $ & $-0.5737$ &    \\
&$\omega \omega$ &  $-0.0100$ & $-0.1463 $ &  \\
\hline\hline
\end{tabular}
\caption{Here we present fall-apart modes into two-vector channels and their relative coupling strengths
from the isovectors, $a_0^+(980)$, $a_0^+(1450)$,
the isodoublets, $K_0^{*+}(800)$, $K_0^{*+}(1430)$, the isoscalars close to the flavor octet, $f_0(500)$, $f_0(1370)$,
and the isoscalars close to the flavor singlet, $f_0(980)$, $f_0(1500)$.
The ratios of the two strengths are also shown.
To get the actual coupling strengths, a common unknown overall factor is needed to be multiplied.
}
\label{couplings}
\end{table}

This trend in the fall-apart strengths into two vector mesons, namely the enhancement for the heavy nonet and the suppression for the light nonet,
is opposite to what we have found for the two-pseudoscalar modes whose fall-apart strengths are suppressed for the heavy nonet
but enhanced for the light nonet~\cite{Kim:2017yur}.
But both are the direct consequences of the tetraquark mixing framework, Eqs.~(\ref{heavy}), (\ref{light}).
So these features are very promising in distinguishing the tetraquark mixing framework from other proposals for the two nonets in the literature.
For the case of the two-pseudoscalar modes, our results
have been tested relatively well for $a_0(980)$, $a_0(1450)$.
That is, our calculated ratios of partial decay widths were found to reproduce successfully the experimental ratios~\cite{Kim:2017yur}.
This leads us to expect that the additional trend in the two-vector modes also exists as an actual phenomenon.

However, unlike to the two-pseudoscalar modes, the two-vector modes are not directly accessible in most
cases due to kinematical constraints.  Most two-vector pairs from the fall-apart modes have invariant masses
above the threshold set by the resonance masses belonging to the two nonets and, therefore,
their partial widths cannot be measured experimentally.
Instead, most coupling strengths presented in Table~\ref{couplings} can play a role of constraints
when one constructs effective Lagrangians involving the participating mesons, which then
can be used to investigate off-shell behaviors of the two nonets.
They can be investigated, for example, in a coupled-channel analysis where the members in the two nonets appear in the intermediate states.

Nevertheless, the two decay modes from the heavy nonet, $f_0(1370)\rightarrow \rho \rho$,  $f_0(1500)\rightarrow \rho \rho$,
should be interesting in comparison with experimental data.
These two modes barely satisfy the kinematical constraints and provide indirect hints for the strength enhancement.
According to PDG, $f_0(1370)\rightarrow \rho \rho$ is reported to be a dominant mode among various modes
in the $f_0(1370)\rightarrow 4\pi$ decays.
Because of the invariant mass,
$2M_{\rho} \sim 1551$ MeV, this decay occurs rarely only from high tails of the broad resonance $f_0(1370)$
and its partial width should be suppressed strongly by the limited phase space.
So its dominance within the $4\pi$ decay modes is difficult to understand unless there is a strong enhancement in the coupling with some
understandable mechanisms.
Indeed, the strong enhancement reported in Table~\ref{couplings} may
provide one possible mechanism for this mode.

We have a similar enhancement for the $f_0(1500)\rightarrow \rho \rho$ mode in Table~\ref{couplings}.
The partial decay width of this mode is also expected to be very small due to the
limited phase space and the fact that
this decay occurs only through the higher tail of the resonance width.
PDG lists this partial decay width
with respect to the $4\pi$ partial width~\cite{Abele:2001pv}, $\Gamma(\rho\rho)/\Gamma(4\pi) \sim 0.13$, even though
this data have been omitted in extracting the resonance parameters in PDG, indicating
perhaps that this measurement needs further confirmation.  Using the branching ratio of $\Gamma(4\pi)/\Gamma_{total}\sim 0.5$,
the data leads to the partial branching ratio, $\Gamma(\rho\rho)/\Gamma_{total}\sim 0.064$.
One can get even larger branching ratio up to 0.4
if one uses the data for $\Gamma(\rho\rho)/\Gamma[2(\pi\pi)_{S-wave}]$ from Ref.~\cite{Barberis:1999wn}
combined with the other ratios given in PDG.  Therefore, although a consensus among the experimental data is lacking,
we see that its branch ratio is not small, which can indirectly support the strong enhancement of the coupling.

Of course, our statement here needs further verification through other reaction mechanisms.
One indirect way is to look into the photoproduction of double $K^0_S$~\cite{Chandavar:2017lgs} through the scalar
resonances $f_0(980)$, $f_0(1500)$ where $f_0 \rho\gamma$ or $f_0\omega\gamma$ vertices
participate in the $t$-channel~\cite{Xing:2018axn}.
Through the vector-meson-dominance, these vertices can be related to $f_0 \rho\rho$ or $f_0\omega\rho$
which can be constrained by our tetraquark mixing framework.

One may ask whether there are experimental supports for the other two-vector modes with similar invariant mass,
$f_0(1370)\rightarrow \omega \omega$,  $f_0(1500)\rightarrow \omega \omega$,
whose couplings are expected to enhance also in our tetraquark mixing model.  Currently in PDG, there is no $\omega\omega$ mode either
from $f_0(1370)$ or $f_0(1500)$.
Apart from the experimental difficulty in measuring $\omega$ in comparison with $\rho$,
one can understand the absence of this mode in two ways.
First of all, the invariant mass of $\omega\omega$, which is 15 MeV larger than $\rho\rho$,
lies in even higher tail of the resonance width.  Also the $\omega$ decay width, $\sim 8$ MeV, is much smaller than the $\rho$ decay width, $\sim 150$ MeV.
So the invariant mass of $\omega\omega$ should be sharp giving it less chance to overlap
with the resonance width of $f_0(1370)$ or $f_0(1500)$.

\section{Summary}
\label{sec:summary}

In this work, we have investigated additional signatures to support tetraquark mixing framework for the two nonets in PDG,
$a_0 (980)$, $K_0^* (800)$, $f_0 (500)$, $f_0 (980)$ in the light nonet,
$a_0 (1450)$, $K_0^* (1430)$, $f_0 (1370)$, $f_0 (1500)$ in the heavy nonet.
In our previous works, the tetraquark mixing framework was tested through
the hyperfine mass splitting generating the experimental mass splitting between the two nonets relatively well.
The predicted fall-apart decay widths into two pseudoscalar mesons is also found to be consistent with the experimental partial widths
as far as the isovector resonances, $a_0 (980)$, $a_0 (1450)$, are concerned.
To solidify this framework, we collect more signatures to identify the two nonets as
the tetraquark nonets in SU(3)$_f$.  These include the Gell-Mann--Okubo mass relation and
the tetraquark mass ordering exhibited from masses of the two nonets.
The marginal mass ordering seen in the heavy nonet could be another signature to support
for the tetraquark mixing framework. Also the tetraquark mixing parameters are found to be independent of isospins
suggesting that the tetraquark mixing framework generates two flavor nonets in SU(3)$_f$ which
can phenomenologically match the flavor structure seen in the two nonets in PDG.
As comparative models, we have examined the two-quark picture with $\ell=1$ and pointed out that
its simple applications are not consistent with the two nonets phenomenologically.
Alternatively the meson-meson bound picture has been discussed also with its possible limitations.
We have emphasized that the fall-apart modes and
their different predictions on the strength between the light and heavy nonets could be a unique feature
to distinguish the tetraquark models from the meson-meson picture and possibly from other models as well.
Indeed, we have calculated the fall-apart coupling strengths into two vector mesons with interesting predictions.
In particular, the coupling strengths of the two-vector modes are found to be enhanced strongly in the heavy nonet while they are suppressed in the light nonet.
Their coupling ratios become huge numbers around $\sim 15$.
This trend in the two-vector modes can provide another testing ground for the tetraquark mixing framework.
We have discussed some experimental hints related to the phenomena particularly from the resonances belonging to the heavy nonet.

\acknowledgments

The work of H. Kim and K.S.Kim was supported by the National Research Foundation of Korea(NRF) grant funded by the
Korea government(MSIT) (No. NRF-2018R1A2B6002432, No. NRF-2018R1A5A1025563).
The work of M.K.Cheoun was supported by the National Research Foundation of Korea (No. NRF-2017R1E1A1A01074023).
The work of D.Jido was supported by the Grant-in-Aid for Scientific Research (No. 17K05449) from JSPS.
The work of M.Oka  was supported by the Grant-in-Aid for Scientific Research (No. 25247036) from JSPS.


\begin{thebibliography}{10}



\bibitem{Choi:2011fc}
  S.-K.~Choi {\it et al.},
  Bounds on the width, mass difference and other properties of $X(3872) \rightarrow \pi^+\pi^- J/\psi$ decays,
  Phys.\ Rev.\ D {\bf 84}, 052004 (2011).



\bibitem{Aaij:2013zoa}
  R.~Aaij {\it et al.} (LHCb Collaboration),
  Determination of the $X(3872)$ meson quantum numbers,
  Phys.\ Rev.\ Lett.\  {\bf 110}, 222001 (2013).


\bibitem{Belle03}
  S.~K. Choi \textit{et~al.} (Belle Collaboration),
  Observation of a narrow charmoniumlike state in exclusive $B^{\pm} \rightarrow K^{\pm} \pi^+ \pi^- J / \psi$ decays,
  Phys.\ Rev.\ Lett.\ {\bf 91}, 262001 (2003).


\bibitem{Aubert:2004zr}
  B.~Aubert {\it et al.} ($BABAR$ Collaboration),
  Search for a charged partner of the $X(3872)$ in the $B$ meson decay $B \to X^- K$, $X^- \to J/\psi \pi^- \pi^0$,
  Phys.\ Rev.\ D {\bf 71}, 031501 (2005).


\bibitem{Aaij:2016iza}
  R.~Aaij {\it et al.} (LHCb Collaboration),
  Observation of $J/\psi\phi$ structures consistent with exotic states from amplitude analysis of $B^+\to J/\psi \phi K^+$ decays,
  Phys.\ Rev.\ Lett.\  {\bf 118}, 022003 (2017).



\bibitem{Aaij:2016nsc}
  R.~Aaij {\it et al.} (LHCb Collaboration),
  Amplitude analysis of $B^+\to J/\psi \phi K^+$ decays,
  Phys.\ Rev.\ D {\bf 95}, 012002 (2017).


\bibitem{Bhardwaj:2013rmw}
  V.~Bhardwaj {\it et al.} (Belle Collaboration),
  Evidence of a new narrow resonance decaying to $\chi_{c1}\gamma$ in $B \to \chi_{c1} \gamma K$,
  Phys.\ Rev.\ Lett.\  {\bf 111}, 032001 (2013).


\bibitem{Xiao:2013iha}
  T.~Xiao, S.~Dobbs, A.~Tomaradze, and K.~K.~Seth,
  Observation of the charged hadron $Z_c^{\pm}(3900)$ and evidence for the neutral $Z_c^0(3900)$ in $e^+e^-\to \pi\pi J/\psi$ at $\sqrt{s}=4170$ MeV,
  Phys.\ Lett.\ B {\bf 727}, 366 (2013).


\bibitem{Abe:2007jna}
  K.~Abe {\it et al.} (Belle Collaboration),
  Observation of a charmoniumlike state produced in association with a $J/\psi$ in $e^+ e^-$ annihilation at $\sqrt{s} \approx 10.6$ GeV,
  Phys.\ Rev.\ Lett.\  {\bf 98}, 082001 (2007).


\bibitem{Aaij:2015tga}
  R.~Aaij {\it et al.} (LHCb Collaboration),
  Observation of $J/\psi p$ resonances consistent with pentaquark states in $\Lambda_b^0 \to J/\psi K^- p$ decays,
  Phys.\ Rev.\ Lett.\  {\bf 115}, 072001 (2015).


\bibitem{D0:2016mwd}
  V.~M.~Abazov {\it et al.} [D0 Collaboration],
  Evidence for a $B_s^0 \pi^\pm$ state,
  Phys.\ Rev.\ Lett.\  {\bf 117}, no. 2, 022003 (2016).


\bibitem{Maiani:2004vq}
  L.~Maiani, F.~Piccinini, A.~D.~Polosa and V.~Riquer,
  Diquark-antidiquarks with hidden or open charm and the nature of $X(3872)$,
  Phys.\ Rev.\ D {\bf 71}, 014028 (2005).

\bibitem{Kim:2016tys}
  Hungchong~Kim, K.~S.~Kim, Myung-Ki~Cheoun, Daisuke~Jido, and Makoto~Oka,
  Testing the tetraquark structure for the $X$ resonances in the low-lying region,
  Eur.\ Phys.\ J.\ A {\bf 52}, no. 7, 184 (2016).


\bibitem{Anwar:2018sol}
  M.~N.~Anwar, J.~Ferretti and E.~Santopinto,
  Spectroscopy of the hidden-charm $[qc][\bar q \bar c]$ and $[sc][\bar s \bar c]$ tetraquarks,
  arXiv:1805.06276 [hep-ph].


\bibitem{Zhao:2014qva}
  L.~Zhao, W.~Z.~Deng and S.~L.~Zhu,
  Hidden-charm tetraquarks and charged $Z_c$ states,
  Phys.\ Rev.\ D {\bf 90}, no. 9, 094031 (2014).

\bibitem{Kim:2014ywa}
  Hungchong~Kim, Myung-Ki~Cheoun, and Yongseok~Oh,
  Four-quark structure of the excited states of heavy mesons,
  Phys.\ Rev.\ D {\bf 91}, 014021 (2015).

\bibitem{Yan:2018gik}
  X.~Yan, B.~Zhong and R.~Zhu,
  Doubly charmed tetraquarks in a diquark-antidiquark model,
  Int.\ J.\ Mod.\ Phys.\ A {\bf 33}, no. 16, 1850096 (2018).


\bibitem{Karliner:2017qjm}
  M.~Karliner and J.~L.~Rosner,
  Discovery of doubly-charmed $\Xi_{cc}$ baryon implies a stable ($b b \bar{u} \bar{d}$) tetraquark,
  Phys.\ Rev.\ Lett.\  {\bf 119}, no. 20, 202001 (2017).


\bibitem{Hyodo:2017hue}
  T.~Hyodo, Y.~R.~Liu, M.~Oka and S.~Yasui,
  Spectroscopy and production of doubly charmed tetraquarks,
  arXiv:1708.05169 [hep-ph].



\bibitem{Esposito:2013fma}
  A.~Esposito, M.~Papinutto, A.~Pilloni, A.~D.~Polosa and N.~Tantalo,
  Doubly charmed tetraquarks in $B_c$ and $\Xi_{bc}$ decays,
  Phys.\ Rev.\ D {\bf 88}, no. 5, 054029 (2013).



\bibitem{Eichten:2017ffp}
  E.~J.~Eichten and C.~Quigg,
  Heavy-quark symmetry implies stable heavy tetraquark mesons $Q_i Q_j \bar q_k \bar q_l$,
  Phys.\ Rev.\ Lett.\  {\bf 119}, no. 20, 202002 (2017).

\bibitem{Chen:2016ont}
  K.~Chen, X.~Liu, J.~Wu, Y.~R.~Liu and S.~L.~Zhu,
  Triply heavy tetraquark states with the $QQ\bar Q \bar q$ configuration,
  Eur.\ Phys.\ J.\ A {\bf 53}, no. 1, 5 (2017).
\bibitem{Lloyd:2003yc}
  R.~J.~Lloyd and J.~P.~Vary,
  All charm tetraquarks,
  Phys.\ Rev.\ D {\bf 70}, 014009 (2004).


\bibitem{Richard:2017vry}
  J.~M.~Richard, A.~Valcarce and J.~Vijande,
  String dynamics and metastability of all-heavy tetraquarks,
  Phys.\ Rev.\ D {\bf 95}, no. 5, 054019 (2017).


\bibitem{Karliner:2016zzc}
  M.~Karliner, S.~Nussinov and J.~L.~Rosner,
  $Q Q \bar Q \bar Q$ states: masses, production, and decays,
  Phys.\ Rev.\ D {\bf 95}, no. 3, 034011 (2017).


\bibitem{Esposito:2018cwh}
  A.~Esposito and A.~D.~Polosa,
  A $bb\bar b\bar b$ di-bottomonium at the LHC,
  Eur.\ Phys.\ J.\ C {\bf 78}, no. 9, 782 (2018).


\bibitem{Jaffe77a}
  R.~L. Jaffe,
  Multiquark hadrons. 1. The Phenomenology of $Q\bar{Q}^2$ mesons,
  Phys.\ Rev.\ D {\bf 15}, 267 (1977).


\bibitem{Jaffe77b}
  R.~L. Jaffe,
  Multiquark hadrons. 2. Methods,
  Phys.\ Rev.\ D {\bf 15}, 281 (1977).


\bibitem{Jaffe04}
  R.~L. Jaffe,
  Exotica,
  Phys.\ Rept.\  {\bf 409}, 1 (2005).



\bibitem{Jaffe:1999ze}
  R.~L.~Jaffe,
  Color, spin, and flavor dependent forces in quantum chromodynamics,
  hep-ph/0001123.

\bibitem{MPPR04a}
  L.~Maiani, F.~Piccinini, A.~D. Polosa, and V.~Riquer,
  A New look at scalar mesons,
  Phys.\ Rev.\ Lett.\  {\bf 93}, 212002 (2004).


\bibitem{EFG09}
  D.~Ebert, R.~Faustov, and V.~Galkin,
  Masses of light tetraquarks and scalar mesons in the relativistic quark model,
  Eur.\ Phys.\ J.\ C {\bf 60}, 273 (2009).

\bibitem{Santopinto:2006my}
  E.~Santopinto and G.~Galata,
  Spectroscopy of tetraquark states,
  Phys.\ Rev.\ C {\bf 75}, 045206 (2007).


\bibitem{Agaev:2017cfz}
  S.~S.~Agaev, K.~Azizi and H.~Sundu,
  The structure, mixing angle, mass and couplings of the light scalar $f_0(500)$ and $f_0(980)$ mesons,
  Phys.\ Lett.\ B {\bf 781}, 279 (2018).



\bibitem{Agaev:2018fvz}
  S.~S.~Agaev, K.~Azizi and H.~Sundu,
  The nonet of the light scalar tetraquarks: the mesons $a_0(980)$ and $K_{0}^{\ast}(800)$,
  arXiv:1804.02519 [hep-ph].

\bibitem{Kim:2016dfq}
  Hungchong~Kim, Myung-Ki~Cheoun, and K.~S.~Kim,
  Spin-1 diquark contributing to the formation of tetraquarks in light mesons,
  Eur.\ Phys.\ J.\ C {\bf 77}, 173 (2017);
  Erratum: Spin-1 diquark contributing to the formation of tetraquarks in light mesons,
  Eur.\ Phys.\ J.\ C {\bf 77}, 545(E) (2017).


\bibitem{Kim:2017yur}
  K.~S.~Kim and Hungchong~Kim,
  Possible signatures for tetraquarks from the decays of $a_0(980)$, $a_0(1450)$,
  Eur.\ Phys.\ J.\ C {\bf 77}, 435 (2017).


\bibitem{Kim:2017yvd}
  Hungchong~Kim, K.~S.~Kim, Myung-Ki~Cheoun and Makoto~Oka,
  Tetraquark mixing framework for isoscalar resonances in light mesons,
  Phys.\ Rev.\ D {\bf 97}, no. 9, 094005 (2018).

\bibitem{Kim:2005gt}
  Hungchong~Kim and Yongseok~Oh,
  $D_s (2317)$ as a four-quark state in QCD sum rules,
  Phys.\ Rev.\ D {\bf 72}, 074012 (2005).


\bibitem{PDG16}
  C.~Patrignani {\it et al.} [Particle Data Group],
  Review of Particle Physics,
  Chin.\ Phys.\ C {\bf 40}, no. 10, 100001 (2016).

\bibitem{Dudek:2016cru}
  J.~J.~Dudek, R.~G.~Edwards, and D.~J.~ Wilson,
  An $a_0$ resonance in strongly coupled $\pi \eta$, $K\overline K$ scattering from lattice QCD,
  Phys.\ Rev.\ D {\bf 93}, 094506 (2016).

\bibitem{Janssen:1994wn}
  G.~Janssen, B.~C.~Pearce, K.~Holinde and J.~Speth,
  On the structure of the scalar mesons $f_0(975)$ and $a_0(980)$,
  Phys.\ Rev.\ D {\bf 52}, 2690 (1995).


\bibitem{Branz:2007xp}
  T.~Branz, T.~Gutsche and V.~E.~Lyubovitskij,
  $f_0(980)$ meson as a $K\overline K$ molecule in a phenomenological Lagrangian approach,
  Eur.\ Phys.\ J.\ A {\bf 37}, 303 (2008).


\bibitem{Branz:2008ha}
  T.~Branz, T.~Gutsche and V.~E.~Lyubovitskij,
  Strong and radiative decays of the scalars $f_0(980)$ and $a_0(980)$ in a hadronic molecule approach,
  Phys.\ Rev.\ D {\bf 78}, 114004 (2008).


\bibitem{Guo:2017jvc}
  F.~K.~Guo, C.~Hanhart, U.~G.~Mei©¬ner, Q.~Wang, Q.~Zhao and B.~S.~Zou,
  Hadronic molecules,
  Rev.\ Mod.\ Phys.\  {\bf 90}, no. 1, 015004 (2018).


\bibitem{Hanhart:2017kbu}
  C.~Hanhart,
  Theory Concepts for Heavy Exotic Mesons,
  Int.\ J.\ Mod.\ Phys.\ Conf.\ Ser.\  {\bf 46}, 1860004 (2018).


\bibitem{Swanson:2004pp}
  E.~S.~Swanson,
  Diagnostic decays of the $X(3872)$,
  Phys.\ Lett.\ B {\bf 598}, 197 (2004).



\bibitem{Tornqvist:2004qy}
  N.~A.~Tornqvist,
  Isospin breaking of the narrow charmonium state of Belle at 3872 MeV as a deuson,
  Phys.\ Lett.\ B {\bf 590}, 209 (2004).


\bibitem{Kim:1995bm}
  Hyun-Chul~Kim and Mikhail~Shmatikov,
  Elusive exotic states,
  Phys.\ Lett.\ B {\bf 375}, 310 (1996).



\bibitem{Wang:2017qbe}
  Q.~Wang,
  Exotic candidates with heavy quark(s),
  PoS Hadron {\bf 2017}, 144 (2018).

\bibitem{vanBeveren:1986ea}
  E.~van Beveren, T.~A.~Rijken, K.~Metzger, C.~Dullemond, G.~Rupp, and J.~E.~Ribeiro,
  A low lying scalar meson nonet in a unitarized meson model,
  Z.\ Phys.\ C {\bf 30}, 615 (1986).

\bibitem{Tornqvist:1995kr}
  N.~A.~Tornqvist,
  Understanding the scalar meson $q\bar q$ nonet,
  Z.\ Phys.\ C {\bf 68}, 647 (1995).



\bibitem{Boglione:2002vv}
  M.~Boglione and M.~R.~Pennington,
  Dynamical generation of scalar mesons,
  Phys.\ Rev.\ D {\bf 65}, 114010 (2002).


\bibitem{Wolkanowski:2015lsa}
  T.~Wolkanowski, F.~Giacosa, and D.~H.~Rischke,
  $a_0(980)$ revisited,
  Phys.\ Rev.\ D {\bf 93}, 014002 (2016).


\bibitem{Maiani:2006rq}
  L.~Maiani, F.~Piccinini, A.~D.~Polosa, and V.~Riquer,
  Positive parity scalar mesons in the 1--2 GeV mass range,
  Eur.\ Phys.\ J.\ C {\bf 50}, 609 (2007).

\bibitem{Black:1998wt}
  D.~Black, A.~H.~Fariborz, F.~Sannino, and J.~Schechter,
  Putative light scalar nonet,
  Phys.\ Rev.\ D {\bf 59}, 074026 (1999).

\bibitem{Black:1999yz}
  D.~Black, A.~H.~Fariborz, and J.~Schechter,
  Mechanism for a next-to-lowest lying scalar meson nonet,
  Phys.\ Rev.\ D {\bf 61}, 074001 (2000).


\bibitem{Dorokhov:1993nw}
  A.~E.~Dorokhov, N.~I.~Kochelev, and Y.~A.~Zubov,
  Four-quark states and nucleon-antinucleon annihilation within the quark model with QCD vacuum-induced interaction,
  Z.\ Phys.\ C {\bf 65}, 667 (1995).



\bibitem{Oh:2004gz}
  Yongseok~Oh and Hungchong~Kim,
  Pentaquark baryons in SU(3) quark model,
  Phys.\ Rev.\ D {\bf 70}, 094022 (2004).




\bibitem{DeRujula:1975qlm}
  A.~De Rujula, H.~Georgi, and S.~L.~Glashow,
  Hadron masses in a gauge theory,
  Phys.\ Rev.\ D {\bf 12}, 147 (1975).


\bibitem{Keren07}
  B.~Keren-Zur,
  Testing confining potentials through meson/baryon hyperfine splitting ratio,
  Annals Phys.\  {\bf 323}, 631 (2008).


\bibitem{Silve92}
  B.~Silvestre-Brac,
  Systematics of $Q^2$($\bar{Q}^2$) systems with a chromomagnetic interaction,
  Phys.\ Rev.\ D {\bf 46}, 2179 (1992).


\bibitem{GR81}
  S.~Gasiorowicz and J.~L. Rosner,
  Hadron spectra and quarks,
  Am.\ J.\ Phys.\  {\bf 49}, 954 (1981).



\bibitem{Sexton:1995kd}
  J.~Sexton, A.~Vaccarino and D.~Weingarten,
  Numerical evidence for the observation of a scalar glueball,
  Phys.\ Rev.\ Lett.\  {\bf 75}, 4563 (1995).


\bibitem{Hyodo:2006yk}
  T.~Hyodo, D.~Jido and A.~Hosaka,
  Exotic hadrons in s-wave chiral dynamics,
  Phys.\ Rev.\ Lett.\  {\bf 97}, 192002 (2006).



\bibitem{Hyodo:2006kg}
  T.~Hyodo, D.~Jido and A.~Hosaka,
  Study of exotic hadrons in s-wave scatterings induced by chiral interaction in the flavor symmetric limit,
  Phys.\ Rev.\ D {\bf 75}, 034002 (2007).


\bibitem{Lee:2004bsa}
  Su~Houng~Lee, Hungchong~Kim, and Yongseok~Oh,
  Decay modes of ideally mixed narrow pentaquark states,
  J.\ Korean Phys.\ Soc.\  {\bf 46}, 774 (2005).


\bibitem{Abele:2001pv}
  A.~Abele {\it et al.} [CRYSTAL BARREL Collaboration],
  $4\pi$ decays of scalar and vector mesons,
  Eur.\ Phys.\ J.\ C {\bf 21}, 261 (2001).



\bibitem{Barberis:1999wn}
  D.~Barberis {\it et al.} [WA102 Collaboration],
  A spin analysis of the $4 \pi$ channels produced in central $pp$ interactions at 450~GeV/c,
  Phys.\ Lett.\ B {\bf 471}, 440 (2000).



\bibitem{Chandavar:2017lgs}
  S.~Chandavar {\it et al.} [CLAS Collaboration],
  Double $K_S^0$ photoproduction off the proton at CLAS,
  Phys.\ Rev.\ C {\bf 97}, no. 2, 025203 (2018).


\bibitem{Xing:2018axn}
  H.~Xing, C.~S.~An, J.~J.~Xie and G.~Li,
  Photoproduction of $f_0(980)$ and $f_0(1500)$ resonances off a proton target,
  arXiv:1807.11151 [hep-ph].

\end{thebibliography}
\end{document}